\input{psfig}
\documentstyle[12pt,aaspp4]{article}

\begin{document}


\title{An Interstellar Conduction Front Within a Wolf-Rayet
Ring Nebula Observed with the GHRS}

 
\author{Bram Boroson\altaffilmark{1}, Richard McCray, 
and Christina Oelfke Clark}
\affil{JILA, Campus Box 440,
University of Colorado,
Boulder, CO 80309; boroson@head-cfa.harvard.edu, dick@jila.colorado.edu,
oelfke@ucsu.colorado.edu}

\author{Jonathan Slavin}
\affil{NASA/Goddard Space Flight Center, Code 685,
Greenbelt, MD 20771; slavin@shock.gsfc.nasa.gov}

\author{Mordecai-Mark Mac Low}
\affil{Max-Planck-Institut f\"ur Astronomie, K\"onigstuhl 17, 69117 
Heidelberg,
Germany; mordecai@mpia-hd.mpg.de}

\author{You-Hua Chu}
\affil{Astronomy Dept., University of Illinois, Urbana,
IL 61801; chu@astro.uiuc.edu}

\and

\author{Dave Van Buren}
\affil{Infrared Processing and Analysis Center,
California Institute of Technology,
Pasadena, CA~91125; dave@ipac.caltech.edu}

\altaffiltext{1}{Center for Astrophysics, 60 Garden Street, Cambridge,
MA 02138}
%
%

\thispagestyle{empty}
\newcommand{\etal}{{et al}.\ }
\newcommand{\lae}{\mathrel{<\kern-1.0em\lower0.9ex\hbox{$\sim$}}}
\newcommand{\gae}{\mathrel{>\kern-1.0em\lower0.9ex\hbox{$\sim$}}}
\newcommand{\Msun}{M_{\odot}}
\newcommand{\Rsun}{R_{\odot}}

\begin{abstract} With the High Resolution Spectrograph aboard the Hubble
Space Telescope we obtained high signal-to-noise ($S/N \gae 200$ --
$600$ per 17 km~s$^{-1}$ resolution element) spectra of narrow
absorption lines toward the Wolf-Rayet star HD~50896.  The ring nebula
S308 that surrounds this star is thought to be caused by a
pressure-driven bubble bounded by circumstellar gas (most likely from
a red supergiant or luminous blue variable progenitor) pushed aside by
a strong stellar wind.  Our observation has shown for the first time
that blueshifted ($\approx$70 km~s$^{-1}$ relative to the star)
absorption components of C\,IV and N\,V arise in a conduction front
between the hot interior of the bubble and the cold shell of swept-up
material.  These lines set limits on models of the conduction front.
Nitrogen in the shell appears to be overabundant by a factor~$~10$.  


The P~Cygni
profiles of N\,V and C\,IV are variable, possibly due to
a suspected binary companion to HD~50896.
\end{abstract}


\section{Introduction}

About 30\% of all Wolf-Rayet (WR) stars are surrounded by ring nebulae
(Lozinskaya 1992).  The strong stellar winds of these stars ($\dot{M} \sim
10^{-5} \Msun$ yr$^{-1}$, $V_{\rm wind} \sim 2000$ km s$^{-1}$) may create
the shells by sweeping up interstellar material (Johnson \&\ Hogg 1965)
or mass lost by the star during previous stages of evolution
(Bisnovatyi-Kogan \&\ Nadezhin 1972; Massevich, Tutukov, \&\ Yungelson
1975). The
theory of expanding interstellar bubbles powered by stellar winds 
(Castor,
McCray \&\ Weaver 1975) correctly predicted that bubbles would be weak X-ray
sources and would show blueshifted absorption lines in their UV spectra.
The dynamical properties of these nebulae have also thus far been roughly
consistent with the theory.  Weaver \etal (1977) refined the theory to
include the nonstationary ionization balance in a radiatively cooling
conduction front between the hot interior of the bubble and the cold
swept-up shell.

The exchange of mass and energy through conductive interfaces helps to
determine 
the global physical structure of the interstellar medium (ISM; McKee \&\ 
Ostriker 1977).  However, 
it is difficult to study these interfaces directly, since we do not know 
their exact locations and physical parameters.  As we show here, a 
Wolf-Rayet ring nebula is an ideal laboratory for studying conductive
interfaces in the ISM.  

There is substantial evidence that in many WR ring nebulae, the
wind does not expand into a uniform medium as assumed (Chu 1983).  For
example, NGC~6888 is elliptical in shape and anomalously faint in X-rays
(Bochkarev 1988; Wrigge, Wendker, \&\ Wisotzki 1994).  RCW~58 contains 
clumps and filaments (Chu 1982; Smith
\etal 1988; Chu 1988), and UV absorption lines are blueshifted depending
on ionization potential, evidence that cold clumps of ejecta are
evaporating in the hot wind (Smith \etal 1984). 

Therefore, to better understand the physics of conduction fronts, we observed
narrow C\,IV and N\,V absorption lines in the spectrum of the bright 
($V\approx6.9$) central WN5 star
HD~50896 (EZ Canis Majoris) of the Sharpless~308 ring nebula.
In contrast to many of these nebulae, the S~308 ring nebula appears 
smooth and nearly spherical (Chu \etal 1982).
The N\,V line should be an excellent diagnostic of the
conduction front, since photoionization by the central star should produce
little of this ion (Esteban \etal 1993). 

We required UV spectra with a high enough signal-to-noise 
ratio to detect the
weak (equivalent width $\sim$5 m\AA) absorption lines that we predicted 
would arise in the conduction front.  With the Goddard High Resolution
Spectrograph (GHRS) aboard the Hubble Space Telescope, we achieved spectra
of this quality.  They show that
the N\,V absorption from the 
S~308 bubble is substantial.

\newpage
\section{Observations}
\label{observe}

The observations took place April 10, 1993 (before the HST servicing 
mission).  We used the
medium resolution G160M grating with the Large Science Aperture (2'' slit),
centered on 1240 \AA\ (for the N\,V doublet) and 1550 \AA\ (for the C\,IV
doublet).  The G160M grating has a spectral range of 35 \AA\ and
resolution of $R \approx 20,000$. 

We expected a count rate high enough that the nonuniformity
(``granularity'') in the photocathode detectors would be a major
source of error (Gilliland \etal 1992, Lambert \etal 1994).  To alleviate
this problem, we used the ``FP-SPLIT'' method, which shifts the spectra 
four
times for each exposure.  In addition, we took three equal subexposures
at each line (though two of the N\,V exposures were at the same central
wavelength).  We used these shifted spectra
to solve for the photocathode granularity (real features would not shift
in wavelength).  Since the sensitivity has an rms variation of 
$\sim$0.5\%\ per diode ($\sim$15 km s$^{-1}$), the error due to
granularity should be $\lae$ 0.2\% merely from the addition of errors,
without considering the solutions for the sensitivity.  Our 
solution for the granularity implied a formal error of $\sim$10$^{-4}$ in 
each diode.

The N\,V line peaked at $1.3 \times 10^{6}$ counts per diode, for a
signal-to-noise (due to counting statistics) ratio of $\sim$10$^{3}$ in 
each 
17 km s$^{-1}$ diode.  The C\,IV line peaked at $2.4 \times 10^{5}$ counts
diode$^{-1}$, for a formal $S/N$ ratio of $\sim$500.  We believe that 
when errors of granularity and counting statistics are included, our 
final $S/N$ ratio is $\sim$200--600 in each diode throughout most of the 
wavelength range observed. 

We recalibrated our wavelength scale by using the ``global wavelength
solution'' (Soderblom, Sherbert \&\ Hulbert 1993).  The additional
wavelength correction implied by the ``Spectrum Y Balance'', or SPYBAL
exposure (Soderblom, Sherbert \&\ Hulbert 1994) was small, approximately 
2~km~s$^{-1}$.  We then derived a wavelength scale for each of our 24
exposures.  We did not, however, combine the spectra according to these
wavelength scales, but correlated features between the spectra.  The
wavelength shifts at which the spectra were best correlated implied an
error of $\sim$7--8 km s$^{-1}$ in the global wavelength solutions. 

Howarth \&\ Schmutz (1995) 
determined the velocity of Na\,I optical absorption lines 
towards HD~50896 from very high resolution 
spectra ($R \approx 10^5$).
  We assume that the absorption from low ionization stages seen in our
spectra (Mg\,II, S\,II, C\,I) should coincide in velocity with the Na\,I
absorption lines. The exposures in the region of the C\,IV lines include
C\,I/C\,I$^{*}$/C\,I$^{**}$ absorption at $\sim$1560 \AA\ at velocities
consistent with the velocity shift of the Na\,I lines.  However, the
absorption lines in the region of 1240 \AA\ are all blueshifted by
$\sim$15 km s$^{-1}$ from the expected velocities.  This finding applies
both to the low-velocity, low-ionization lines (Mg\,II and S\,II), which
are shifted relative to the Na\,I lines, and also to the N\,V absorption
(which we attribute to the interstellar bubble S~308),
which is shifted by $\sim$15 km s$^{-1}$ from the [O\,III] emission seen
by Chu \etal (1982).  There is a 3 km s$^{-1}$ difference between the
S\,II and Mg\,II velocities; we have set our calibration so that the Na\,I
absorption would fall between the Mg\,II and S\,II velocities. 

Although the calibration shift of 15 km s$^{-1}$ is greater than the error 
expected from the variability among our separate wavelength solutions, 
the resulting agreement of velocities among all low-ionization species 
and between the N\,V absorption and the [O\,III] emission 
leads us to add 15 km 
s$^{-1}$ to all velocities in our exposure in the range of 1240 \AA.

The exposures centered on 1240 \AA\ lasted 27.2 min., whereas the 
1550~\AA\ exposures lasted 16.3 min.  The spectra (shown in Fig.~1) are
of much higher quality than those obtained with Copernicus (Shull 1977)
and IUE (Howarth \&\ Phillips 1986; Nichols-Bohlin \&\ Fesen 1986; Smith,
Willis, \&\ Wilson 1980).  The spectra of Howarth \&\ Phillips (1986), 
with a signal to noise ratio of $\sim$50--75, were the best prior to our 
observation.  In addition to the 
absorption lines of N\,V and C\,IV, we also
see Mg\,II$\lambda\lambda$1239,1240, S\,II$\lambda\lambda$1250,1253, and
C\,I/C\,I$^{*}$/C\,I$^{**}\lambda\lambda\lambda$1560,1560.7,1563.
Equivalent widths and Local
Standard of Rest velocities are listed in Table~1 for low-velocity
features (and blended features), and in Table~2 for high-velocity features
($V_{\rm LSR} = V_{\rm hel} - 19.5$ km s$^{-1}$). 

Tables 1 and 2 also display results from the literature 
for a variety of UV absorption lines.
Use of both LSR and heliocentric velocities has made
comparisons of published values difficult. Also, absolute wavelength
calibration has not been reliable; for this reason, some papers list all
velocities relative to the low-velocity component.  Following the
suggestion of Howarth \&\ Schmutz (1995), we have added +39 km s$^{-1}$ to
the IUE-determined velocities of Howarth \&\ Phillips (1986) to convert
them to heliocentric velocities.  We recalibrated with recent and
accurate rest wavelength determinations (Verner, Barthel, \&\ Tytler
1994).

By employing deconvolution techniques (Lucy 1974; Jansson 1984), we
can improve spectral resolution at the cost of increasing the noise. In
Figure~2 we display spectra deconvolved by using the slightly superior
Jansson technique applied to data presmoothed with a Wiener filter.

The C\,IV and N\,V lines are doublets.  Since the blue components have
twice the oscillator strength of the red components, the optical depths in
the absorption lines should also have the expected 2:1 ratio for all 
velocities.
Figure~3 shows the optical depth as a function of velocity, $\tau(v)$, in
the deconvolved spectra.  The two components $\tau_{\rm
blue}(v)$ and $2 \tau_{\rm red}(v)$ can be compared to give a rough
estimate of our total error.  To find $\tau$, we fit several overlapping
parts of the spectrum with polynomials, excluding the regions
contaminated by lines.  We assume that these polynomials represent the
unabsorbed continuum.  Where the absorption is saturated, we cannot
accurately measure $\tau$, and here the less saturated line $\tau_{\rm 
red}$ provides a better estimate.  While the N\,V and Mg\,II lines are 
shallow and unsaturated, the S\,II and C\,IV lines become saturated toward 
line center, and depart from a 2:1 ratio of $\tau_{\rm blue}$ to 
$\tau_{\rm red}$.  We suspect that the saturation of these lines leads to an 
underestimate of the optical depths, especially in the blue components.

Figure 4 shows the ratio of the optical depths in the C\,IV and N\,V
lines.  Near the centers of the lines, the ratios for the two doublet
components disagree because of the saturation of C\,IV.  (The red
components suffer less from saturation and are probably more accurate.) If
we had not made the 15 km s$^{-1}$ shift in the exposure containing the
N\,V lines, then N\,V would be overabundant at a blue shift of $\sim$60 km
s$^{-1}$ LSR. 

Figure 5 shows a schematic view of the velocities of the 
absorption lines observed with the {\it GHRS}, the velocity shifts we have 
made, and the velocities attributed to HD~50896, the expanding shell of 
S~308, and a putative supernova remnant (Nichols-Bohlin \&\ Fesen 1986)
from observations at other wavelengths.


\section{Results and Interpretations}

\subsection{The Conduction Front in S 308}
\label{ism}

Because radiative cooling peaks at a temperature $T\approx10^5$K,
interstellar gas is expected to be predominantly hot ($T\approx
10^7-10^8$K, McCray 1987) or cool ($T\approx10^4$K).  At the interfaces 
between
interstellar gas at these stable temperatures, thermal conduction will 
evaporate cold material into a warm transition region of $T\approx10^5$~K, 
ideal for the formation of O\,VI, N\,V, and C\,IV (Cowie \&\ McKee 1977;
McKee \&\ Cowie 1977; Balbus \&\ McKee 1982).  Observations of absorption 
due to these ions is not conclusive evidence of a conduction front, however,
since photionization and shocks can also be present.  In
the case of S~308, we know the velocity of the expanding shell from
optical observations (Chu \etal 1982) and recent ROSAT observations
(Wrigge 1996) show the presence of hot gas in the interior.  Thus
we have a strong expectation for finding a conduction front along the
line of sight towards HD~50896.

The argument that the N\,V absorption arises in a conduction front is
bolstered by a comparison of the observed N\,V absorption with absorption
towards other stars with sight-lines that do not intersect known
bubbles.  From observations with the {\it GHRS} of interstellar absorption 
towards 12 stars (Sembach \&\ Savage 1992), we would expect
$N(N\,V)\sim 4\times10^{12}$ cm$^{-2}$ (assuming for HD~50896 a
distance of 2~kpc and a height above the galactic plane of $z\sim$300
pc).  This column density is similar to that observed for the
low-velocity N\,V gas ($N(N\,V)\approx5\times10^{12}$ cm$^{-2}$,
see Table~4), but an order of magnitude below that observed
for the high-velocity gas.
We also note that from the model of interstellar absorption
of Sembach \&\ Savage, we would expect a C\,IV column density
of $N(C\,IV)\sim10^{13}$ cm$^{-2}$, a factor of $\sim15$ below
that observed towards HD~50896.

Whether the ring nebula surrounding HD~50896 results from a wind
interacting with stellar ejecta or interstellar material, a conduction
front should form.  
To model the column densities of C\,IV and N\,V in either case, we used 
the analytic approximations given by Weaver \etal (1977), scaled
according to the abundances appropriate for S~308.
That is, the column density of O\,VI, $N({\rm O\,VI})$, is given by
\begin{equation}
N({\rm O\,VI})\approx5.2\times10^{16}X_{\rm 
O}n_0^{9/35}L_{36}^{1/35}t_6^{8/35}{\rm cm}^{-2},
\end{equation}
where $X_{\rm O}$ (given by Weaver \etal as 4.4$\times10^{-4}$) is the 
fractional abundance of oxygen, $n_0$ is 
the ambient proton density, $L_{36}$ is the wind luminosity in units of 
$10^{36}$~erg~s$^{-1}$ ($L\equiv 1/2 \dot{M} V_{\rm w}^2$),
and $t_6$ is 
the age of the bubble in $10^6$ years.  These variables can be reduced
to observed quantities through
\begin{equation}
\rho_0= \frac{125}{154 \pi} t^3 L / R_{\rm s}^5
\end{equation}
and
\begin{equation}
t=(3/5) R_{\rm s} / V_{\rm shell}
\end{equation}
where $\rho_0$ is the mass density, $R_{\rm s}$ is the radius of the shell,
and $V_{\rm shell}$ is its velocity.  To find column densities of nitrogen 
and carbon, we use $N(N\,V)/N(O\,VI)=-1.2$ and $N(C\,IV)/N(O\,VI)=-0.8$,
as given in Weaver \etal (1977).  We then scaled the resulting column 
densities according to the abundances given in Table~4.

We find that, in spite of our incomplete
knowledge of S~308 and HD~50896, the theory predicts a narrow range of
possible column densities for these ions (Table~4).  In Table~4, we also
display the column densities calculated from the observed equivalent
widths (shown in Table~1 and Table~2).  We assume the N\,V lines are on
the linear portion of the curve of growth (the equivalent width is
proportional to the column density), while we obtain a lower limit for the
column density of the C\,IV lines (which probably contain several
components) using the deconvolved line profiles.  
The saturated C\,IV absorption lines are far too strong to arise in a 
conduction front, but probably originate either in another ionization front
or as a result of photoionization.

The strong blueshifted absorption component of N\,V has most recently been
attributed to an unobserved supernova remnant along the line of sight to
HD~50896 (Howarth \&\ Phillips 1986; Nichols-Bohlin \&\ Fesen 1986).  IUE
observations of stars within a region of $3^{\circ}$ around HD~50896 show
blueshifted absorption from low-ionization species at high velocities 
(Nichols-Bohlin \&\ Fesen 1986;
Howarth \&\ Phillips 1986). However, the IUE spectra of stars in this
region did not have a high enough S/N to detect high velocity absorption
from species such as C\,IV, Si\,IV, and N\,V at column densities similar
to those observed in HD~50896.
Furthermore, redshifted absorption corresponding to the receding
hemisphere of the shell has not been observed, although this does not rule
out an origin in a SNR (Nichols-Bohlin \&\ Fesen 1986). 

Both the putative SNR and the ring nebula S~308 expand at velocities near
that of the N\,V absorber.  We identify this absorption with the ring
nebula and not the SNR because the SNR is not observed at other 
wavelengths, and only absorption from low ionization species has been 
seen near $-50$ km s$^{-1}$~LSR from nearby stars.

Without proper calibration for the IUE velocities, 
Howarth \&\ Phillips (1986) identified 
the shallow, low-velocity N\,V absorption at
$\approx$20 km s$^{-1}$ LSR with S~308.  However, this velocity
is inconsistent with the 
emission velocity of
$-40$ km s$^{-1}$ LSR seen by Chu \etal (1982) from several lines of sight to
S~308.  In our model, the strong absorption component arises in a shell
expanding at 80 km s$^{-1}$ (the hot gas expands $\sim$5 km s$^{-1}$
slower than the cold edge of the shell).  Since the central star has a
velocity of +33 km s$^{-1}$ LSR (Chu \etal 1982), we expect to see N\,V
absorption at $-$42 km s$^{-1}$.  The actual N\,V absorption appears at $-$50
km s$^{-1}$; this velocity difference could be attributed to calibration
errors (the alignment of Mg\,II and S\,II with the Na\,I interstellar
velocity) or the nitrogen could be moving faster than the [O\,III]
observed by Chu \etal (1982). 

The deconvolved S\,II line profile (Figure~3d) shows weak absorption at $-$50
km~s$^{-1}$ LSR, consistent with an origin in the the same gas as the N\,V
absorber.  The equivalent width of the weak S\,II line implies an
absorbing column density of $\log(N)=13.9\pm0.1$ cm$^{-2}$ (Howarth \&\ 
Phillips give $\log(N)=13.51\pm0.21$).  We 
only see this line in our deconvolved spectra, and the Jansson 
deconvolution technique may lead to an overestimate of the column
density of a weak absorption feature because it forces the flux to be less
than the continuum level.  The presence of S\,II at the same velocity as N\,V
suggests that the gas at $-50$ km~s$^{-1}$ LSR
contains several temperature zones.  This would be expected to occur
in both a SNR shock front and in the
conductive interface between a cold shell (containing S\,II) and the hot 
interior of S~308. 

We considered a range of models to fit the data (Table~4).
We adopt Model~E as the most likely set of parameters for HD~50896/S~308.
This model assumes that HD~50896 is distant ($D=2$ kpc) and that the shell
expansion is fast (80 km~s$^{-1}$), and we use the wind speed at which the 
P~Cygni absorption is saturated (1700 km~s$^{-1}$).  The N\,V column
density predicted by this model is about a factor of~10 below the observed
value.  Thus S~308 is probably enriched in nitrogen, as would be expected if 
the shell is swept-up circumstellar ejecta.  
While our models~A and~B agree well with the
observed N\,V column density, we have rejected the identification
of the low-velocity absorption with the conduction front because it
does not match the expected velocity.  We therefore identify the
high-velocity N\,V 
absorption with a nitrogen-enriched conduction front.

\subsection{P~Cygni Lines and Variability}


The P~Cygni line profiles are consistent with the previously reported wind
velocity of 1700 km~s$^{-1}$ (this is the edge of the saturated
absorption; the absorption edge at 2700 km~s$^{-1}$ is not typical of most
of the wind material).  Table~3 contains information on the P~Cygni lines
during each FP-SPLIT subexposure.  The redshifted part of the N\,V P~Cygni
profile varied by as much as 5\% within 5 min.  It would be unusual for a
variation this strong to result from calibration problems.  If the
aperture projection onto the photocathode relative to the diodes drifts
vertically due to geomagnetic interference, one might expect to see the
spectrum to change with a gradient with wavelength (Gilliland 1994). We
examined the change in the spectrum over a 10~minute 
interval and found that at wavelengths longer than the saturated P~Cygni 
absorption, there was a roughly constant fractional decrease in the flux, 
while there was a roughly constant fractional increase in the flux
at shorter wavelengths.  While this might suggest an 
instrumental problem,
it is curious that the spectrum should pivot about a wavelength
in the region of saturated P~Cygni absorption.

Assuming that the spectral variation is not instrumental,
it could be the result of a compact companion
to HD\,50896 suggested by photometry and polarimetry (a
3.77-day period was found by Drissen \etal 1989, and a possible 0.11
second period was suggested by Marchenko \etal 1994).

We have modeled the N\,V profile by assuming spherical symmetry and
integrating the Sobolev source function exactly (using 
the SEI method of Lamers \etal 1987, which also allows wind turbulence).  A
variable X-ray source with X-ray luminosity $L_{\rm X} \sim 10^{36}$ 
erg~s$^{-1}$ embedded in a stellar wind can produce rapid changes in the
P~Cygni profiles of ions such as C\,IV and N\,V (Kallman, McCray, \&\ Voit
1987; Boroson \etal 1996).  However, this model may not apply to HD~50896,
for which $L_{\rm X} \approx 10^{33}$ ergs s$^{-1}$ (Pollock 1987) and the
red parts of the N\,V and C\,IV profiles are too strong to result entirely
from resonance scattering.  The emission is provided instead by
collisional excitations, which are sensitive to temperature and
velocity variations with distance from HD~50896.  
A small uniform change in the wind temperature (from 60,000 K to 58,500 K)
could account for the observed change in the P~Cygni line equivalent width. 

Though the exact mechanism for the rapid P~Cygni profile variability is 
not yet understood, only a small perturbation of the wind is needed to 
account for it. 

\subsection{The Energy Budget Problem of S~308}

Van Buren (1986) pointed out that the kinetic energy of the shell S~308 is
a fraction $\epsilon_{\rm s}\approx0.05$ of the energy deposited by the
stellar wind, whereas the theory developed by Weaver \etal (1977) predicts
$\epsilon_{\rm s}=0.2$.  This discrepancy could be resolved if the shell
is more massive than Van Buren supposed ($M_{\rm s}=220 \Msun$), but this
would imply ambient interstellar densities of $n_0>4$ cm$^{-3}$. This is
unlikely because HD 50896 is $\sim$300 pc above the galactic plane and may
be in the low-density interior of a bubble swept out during earlier stages
of the star's evolution. 

If S~308 is not of interstellar origin but contains gas supplied by the
star itself during a previous stage of mass loss (Garcia-Segura \&\ Mac
Low 1995) then the mass implied for S308 by energy considerations ($\sim 400
\Msun$) is more than can reasonably have been supplied by HD~50896 and its
possible companion. 

The energy budget problem could be solved if the WR wind had a much lower
mass-loss rate (several $10^{-6}\Msun$~yr$^{-1}$) during the dynamical age
of the shell ($\approx10^5$ years).  Indeed, determinations of current 
mass-loss rates
for WR stars from radio observations may be in error by a
factor of $\gae3$ if the wind is clumpy (Moffat \&\ Robert 1994).  We have
theoretical expectations that the mass-loss rates of WR stars are
variable (Garcia-Segura, Mc Low, \&\ Langer 1996).  The kinetic energy
of S~308 should be properly compared not with the current wind luminosity,
but with the wind luminosity over the dynamical age of the bubble,
which may be much less.
In this case, our model predictions of the N\,V column density in the 
conduction front (Table~4) should not need to be revised,
since the bubble theory (Weaver \etal 1977) predicts 
$N(N\,V)\sim L^{1/35}$. 

We note that a similar energy budget problem for the WR ring nebula
NGC~6888 (Garcia-Segura \&\ Mac Low 1995) could be resolved if the WR mass
loss rate during the formation of NGC~6888 was also much lower than that
inferred at the present epoch. 

\subsection{Density inferred from C\,I line ratios}

The low-velocity lines of C\,I,C\,I$^{*}$,C\,I$^{**}$ 
($\lambda=1560,1560.7,1563$) provide a useful 
density diagnostic.  The neutral absorption towards
HD~50896 has been investigated by Howarth \&\ Schmutz (1995),
who resolved separate narrow ($<3$ km~s$^{-1}$) components of Na\,I.
From their analysis, we expect the C\,I absorption to be
saturated.  In this case, the ratios of the equivalent widths
of the C\,I lines will not suffice to determine electron
density.

To estimate a density, we decompose the
C\,I lines into the same components as the Na\,I low-velocity
lines observed by Howarth \&\ Schmutz (1995).  Since our data only provide
a resolution $R \approx 20,000$, we cannot expect to deduce the density in 
each
individual absorption component; instead we assume that every absorber has
the same electron density $n_{\rm e}$.

We model the three C\,I lines by assuming further that each of the ratios
Na\,I/C\,I, 
Na\,I/C\,I$^{*}$, and Na\,I/C\,I$^{**}$ has the same value in every absorber.
The 90\% confidence range in these ratios thus determines the 
likely range of electron density (there is a weak dependence on 
temperature and ionization state of the gas).  Figure~6 shows the 
observed C\,I lines superimposed on the model derived from the Na\,I lines.
We also show the line profile expected after the HRS line spread function is 
applied.

We include the effects of collisional excitation of C\,I by protons
(Roueff \&\ Bourlot 1990) and neutral hydrogen (Launay \&\ Roueff 1977).
For electron impact excitation, we use the results of Johnson, Burke, \&\
Kingston (1987). 

The results are shown in Figure 7.  Clearly, the values of $n_{\rm e}$ 
implied by 
the separate line ratios (C\,I$^{*}$/C\,I, C\,I$^{**}$/C\,I, and 
C\,I$^{*}$/C\,I$^{**}$) are consistent for a wide range of temperatures and 
degrees of ionization.  If the gas containing the C\,I is largely 
neutral, then $n_{\rm e} \gae 100$ cm$^{-3}$, whereas if the gas is ionized, 
then $n_{\rm e} \sim 1$ cm$^{-3}$.  Inferred densities are lower if the 
gas is hotter.

\subsection{Low-velocity Absorption Components}

Two observing campaigns have targeted stars in the field of view of
HD~50896; Howarth \&\ Phillips (1986) used IUE, and Howarth \&\ Schmutz
(1995) observed Na\,I lines at high resolution.  We note that none of the
stars observed with IUE showed C\,IV absorption comparable in equivalent
width to that observed towards HD~50896.  Among the stars observed
by Howarth \&\ Schmutz (1995), only HD~50562 and HD~51854 show absorption
as complex as HD~50896, including absorption at $\gae$30 km~s$^{-1}$ LSR
(we exclude the star $-$22 1661, which is farther away at $D=3.4$ kpc).
It is likely that the C\,IV absorption arises in gas surrounding
HD~50896.  The C\,IV absorption is redshifted by $\approx 18$ km~s$^{-1}$
from HD~50896; this may represent the expansion speed of a red supergiant 
or luminous blue variable wind which is being swept up by the bubble and 
photoionized by the central star.

We note that the Na\,I spectra of Howarth \&\ Schmutz show 
absorption at $+14.5$ km~s$^{-1}$ LSR
which is seen only in HD~50896 and not in nearby 
stars.  This absorption component
may be associated with HD~50896; perhaps not with the nebula S~308 (which
expands at a higher velocity), but with the wind of a red supergiant 
progenitor.  The problem with such an interpretation is that
the WR~star is likely to ionize the circumstellar gas.  While
the EUV spectrum of HD~50896 and the neutral mass of the shell
are both uncertain, Chu \etal (1982) have argued that the nearby H~II 
regions S~303 and S~304 are photoionized by HD~50896.  We have also just 
attributed the C\,IV absorption to photoionized gas surrounding S~308.  
From this it would follow 
that S~308 is density bounded and that circumstellar Na\,I could only 
exist in dense clumps.  There may be dense neutral clumps of compressed red 
supergiant wind within S~308, but it is also possible that
the structure of neutral absorbers is complex at small scales and
the $+14.5$~km~s$^{-1}$~LSR component arises elsewhere.

The low-velocity absorption probably contains unresolved 
components that arise in physically distinct sources.  The relative 
strengths of these absorption components may be different for each atomic 
absorber.  For example, the strong 
low-velocity C\,II absorption seen with IUE is seen in other nearby 
stars, whereas strong low-velocity C\,IV absorption is seen only toward
HD~50896.  It is also possible
that the velocities of the Mg\,II and S\,II absorption lines, which we have 
used to 
calibrate our observation in the region of N\,V, do {\em not} match the 
velocity of the averaged Na\,I lines.  This mismatch could result, for 
example, if one absorption 
component is enriched in Mg\,II or S\,II, from depletion in grains, or 
as a result of the different ionization potentials.

\subsection{Summary and Future Work}

These short exposures with HST (pre-COSTAR) provided spectra with $S/N$ an
order of magnitude greater than all the averaged IUE spectra (Howarth \&\
Phillips 1986).  We identified absorption lines from the conduction front, 
compared these absorption lines with the predictions of Weaver \etal (1977), 
and inferred that the gas is enriched in nitrogen.  In addition, we saw 
variability in
the wind of HD~50896 on a very short timescale ($<$5 min.).  

The dynamical evolution of a bubble depends on the amount of energy radiated
away in the conduction front.  The N\,V ion may be an important 
contributor to the energy loss of the bubble.  Our observation
that this ion is present (and enhanced by a factor of $\sim10$)
may bear on the question of the future evolution of the bubble.
All our models meet the limit set on O\,VI absorption by 
{\it Copernicus} (Shull 1977); even more energy is likely to be
radiated away by O\,VI so future observations of this absorption
feature (with FUSE, for example) are likely to be extremely important.  

Our observations will constrain models of conductive interfaces that 
include physical effects we have neglected, such as
the saturation of conduction (Cowie \&\ McKee 
1977) and the evolution of instabilities in the shell 
(Garcia-Segura \&\ Mac Low 1994).

Optical observations could determine the density in the shell, and 
thus its mass.  Such data would show whether S~308 is too massive to be of 
stellar origin.
Observations of conduction fronts in similar structures could
indicate whether nitrogen enhancement is a common phenomenon.
Finally, observations using the Space Telescope Imaging Spectrograph (STIS) 
should greatly enhance our understanding of ring nebulae around WR
stars and conduction fronts in the interstellar medium.

\vspace{2 em}

For their
advice and assistance, we would like to thank Ron
Gilliland, Linda Sherbert, Lauretta Nagel, Rob Fesen, and Ian Howarth.

This work was based on observations with the NASA/ESA Hubble Space 
Telescope, obtained at the Space Telescope Science Institute, which is 
operated by the Association of Universities for Research in Astronomy, 
Inc., under NASA contract NAS 5-26555.

\newpage

\section*{TABLES}
\scriptsize
\begin{deluxetable}{lllll}
\tablecaption{Low-velocity Lines in HD\,50896}
\tablewidth{0pt}
\tablehead{
\colhead{Line} & \colhead{$\lambda$\tablenotemark{a}} & \colhead{$v$ (LSR) } &
\colhead{EW(m\AA)} & \colhead{Ref\tablenotemark{b}}}
\startdata
C\,I &  1188.833 &  7.7 &  11.6 &  1\\
 &      1188.833 &  -2.6 &  5.6 &   1\\
 &      1188.833 &  3.5 &   20$\pm$10 &    2\\
 &      1193.030 &  25.5 &    9$\pm$8 &     2\\
 &      1260.736 &  7.5 &  38.7  & 1\\
 &      1260.736 &  6.1 &  13 &    1\\
 &      1260.736 &  22.8 &   33$\pm$5 &    2\\
 &      1260.736 &   7.8 &   37 &      1\\
 &      1261 blend* &  16 &   5$\pm$ 3&      2\\
 &      1260.736 &  26.2&   6&      2\\
 &      1276.482 &  16.7 & 7$\pm$5 & 2\\
 &      1276.482&   10.8&  97.6&   1\\
 &      1277.245&   22.5 &   85$\pm$7&     2\\
 &      1277.513* &   25.5&     24$\pm$4&     2\\
 &      1279.890* & 24.2 & 12$\pm$4 & 2\\
 &      1280.135&   27.7&   28$\pm$5&     2\\
 &  	1328.833&   21.0 & 63$\pm$5 & 2\\
 &      1329.101*&   26.0&   37$\pm$4&     2\\
 &      1329.584**&   20.9&   10$\pm$3&     2\\
 &      1560.3092&   20.5 &      110&    *\\
 &      1560.3092&   21.5&     102$\pm$9&    2\\
 &      1560.69*&   22.0 &      54&     *\\
 &      1560.69*&   40.5&    55$\pm$6&     2\\
 &      1561.438*&   21.5&   10$\pm$4&     2\\
 &      1561.438*&   20.5 &      27&     *\\
 &      1656.928&   21.5&     172$\pm$13&    2\\
 &      1656.266&   43.5 &   47$\pm$7 &    2\\

C\,II  &      1334.532&   14.5,-62,-115 &     556$\pm$35&      2\\
 &      1335.703&   16.5,-57.5,&     191$\pm$12&    2\\

C\,IV & 1548.195&   15& 419$\pm4$&    *\\
 &      1548.195&   15&     447$\pm$31&      2\\
 &      1550.770&   15& 337$\pm26$&    *\\
 &      1550.770&   17.5&     328$\pm$3&    2\\
\tablebreak
N\,I &	1134 &	+2.9 &	165 &	1\\
 &	1135 &	+2.1 &	187 &	1\\
 &	1200 &	+1.4 &	205 &	1\\
 &	1200 &	+10 &	198 &	2\\
 &	1200 &	18.5 &	207 &	2\\
 &	1201 &	14.5 &	174 &	2\\

N\,II &	1084 &	+.3 &	198 &	1\\
 &	1086 &	-.6 &	11.0 &	1\\
 &	1086 &	-1.4 &	13.2 &	1\\

N\,V &  1238.821 & +25 & 14 & * \\
     &  1238.821 & +27.5 & 7$\pm6$& 2 \\
&       1242.804 & +25 &        5 &     * \\
O\,I &  1039&   +4.3&   179&    1\\
 &      1302&   15.5&     229&    2\\
O\,VI & 1038 &  ? &     $<$44 &   1\\

Na\,I & 5890,5896 &  4.8 &    &   3\\
      & 5890,5896 &  8.5 &   &  3\\
 &      5890,5896 & 14.5 &  &   3\\
 &      5890,5896 & 18.0 &  &  3\\ 
 &      5890,5896 & 22.2 &  &  3\\
 &      5890,5896 & 27.1 &  &  3\\
 &      5890,5896 & 29.5 &  &  3\\
 &      5890,5896 & 32.5 &  &  3\\

Mg\,II
&       1239.925&   18.3&    33$\pm5$&     *\\
&       1240.395&   18.8&    17$\pm2$&     *\\
&       1239.925&   18.3&   35$\pm$9&     2\\
&       1240.395& 13.5&     18$\pm$8&     2\\

Al\,II &        1670.787&   18,-68,-127&    335$\pm$23&    2\\

Al\,III &      1854.716&    29.5 &     129$\pm$ 11&    2\\
 &      1862.790&    16.5&     94$\pm$10 & 2\\
\tablebreak

Si\,II&	1021&   ?&	168&	1\\
 &	1190	&-.2&	200&	1\\
 & 	1193&	+1.5&	242&	1\\
 &	1304.372 &	21.4&	174$\pm$12&	2\\
 &	1526.708&	25.5&	237$\pm$17&	2\\
 &	1808.021&	24.1&	143$\pm$11&	2\\

Si\,III &1206.51&	-5.9&	223&	1\\
 &	1206.51&	9.5,-65,-112&	398$\pm12$&	2\\

Si\,IV &1393.755&	19,-50.4&	398$\pm12$&	2\\
 &	1402.770&	26,-45&	167$\pm13$&	2\\

P\,II&	1152.81&	22&	121$\pm27$&	2\\
 &	1532.51&	19.5&	7&	2\\

P\,III &	1334.866&	13.5&	13&	2\\

S\,I &	1807.311&	32.6&	9$\pm 5$&	2\\

S\,II &	1250.578&	 1.3&	146&	1\\
 &	1250.578&	21.4&	134$\pm10$&	2\\
 &	1250.578&	21.5&	141$\pm3$ &	*\\
 &	1253.805&	+1.4&	174&	1\\
 &	1253.805&	23.1&	140$\pm10$&	2\\
 &	1253.805&	 21.9&	173$\pm4$ &	*\\
 &	1259.518&	0.6&	185&	1\\
 &	1259.518&	27&	154&	2\\

S\,III & 1013&	+1.4&	104&	1\\
 &	1190&	+.1&	96&	1\\

S\,IV & 1063&	+8.1&	45&	1\\

Cl\,I &	1347.24&	23.5&	19$\pm4$	&2\\

Cl\,II & 1071&	+.3&	13.9&	1\\

Ar\,I &	1048&	+5.9&	156&	1\\
 &	1067&	+4&	144&	1\\
\tablebreak

Ca\,II & 3933.663&	33&	200$\pm15$&	2\\
 &	3968.468&	33.3&	120$\pm40$&	2\\

V\,III &	1153.179&	6.2&	35$\pm26$&	2\\
 &	1169.262&	47&	15$\pm10$&	2\\

Mn\,II & 1197.184&	11&	22$\pm6$&	2\\
 &	1201.118&	17.5&	10$\pm5$&	2\\

Fe\,II &        1055&   +1.7&   37 &    1\\
&       1097&   -1.6&   87&     1\\
&       1145&   0.0&    151&    1\\

Ni\,II &	1317.217&	21&	25$\pm4$ &	2\\
&	1370.132&	25.6&	26$\pm5$&	2\\
&	1454.842&	15&	7$\pm4$&	2\\
&	1741.549&	18&	24$\pm4$&	2\\
&	1751&	12.6&	24$\pm6$&	2\\

Ge\,II &	1237.0589 &	17 &	25$\pm9$&	4\\

\enddata
\tablenotetext{a}{An asterisk in the ``$\lambda$'' column indicates absorption from an excited 
level.}
\tablenotetext{b}{
References: 1---Shull (1977),
2---Howarth \&\ Phillips (1986),
3---Howarth \&\ Schmutz (1995),
*---This work}
\end{deluxetable}
\normalsize

\newpage

\scriptsize
\begin{deluxetable}{lllll}
\tablecaption{High-velocity lines in HD\,50896}
\tablehead{
\colhead{Line} & \colhead{$\lambda$ \tablenotemark{a}} & \colhead{$v$ (LSR)} &
\colhead{EW (m\AA)} & \colhead{Ref\tablenotemark{b}}}

\startdata
C\,II  &       1334.532 &   -62,-115 & 556 &    2 \\
C\,II$^{*}$ &       1335.703 &   -58   &  191  &   2 \\

C\,IV &     1548.195  &  -21-110 &447$\pm21$ &    2\\
  &      1550.770   & -30-111& 337   &  2\\      

N\,I&	1134.66 &	? &	$<$5.8&	1\\
&	1134.66 &	-73.3&	17.2&	1\\
&	1199.55 &	-65&	60$\pm7$&	2\\
&	1200.223 &	-61&	43$\pm6$&	2\\
&	1200&	-75.1&	44&	1\\		
&	1200&	-76.9&	56&	1\\	
N\,II&	1083.994&	-76.3&	85.2&	1\\
&	1083.994&	-77.7&	84.2&	1\\
&	1083.994&	-84.1&	108&	1\\
N\,II$^{*}$&	1084.575&	-81&	0.9&	1\\
N\,V&	1238.821&	-53&	44$\pm9$&	2\\
     &  1238.821 & -50, total blend   &   66$\pm3$ & * \\
&	1242.804&	-61&	22$\pm7$&	2\\
&       1242.804&  -50, total blend   &   32$\pm2$ & * \\ 
O\,I &      1039.23 &   -73.9&   58.6 &   1\\
&        1302.1685 &   -77.9 &  100  &   1\\
&       1302.1685 &   -63  &   125$\pm9$  &   2\\
Al\,II&    1670.7874 &   -81&     76    &  1\\
     &   1670.7874&    -68-127& 335$\pm 23$ &     2\\

\tablebreak

Si\,II&	1190.4158&	-71.6&	113&	1\\
&	1190.4158&	-9-52&	350$\pm26$&	2\\	
&	1193.2897&	-71.6&	137&	1\\
&	1193.2897&	-72.6&	151&	1\\
&	1193.2897&	-74.6&	139&	1\\
&	1193.2897&	+19.5-56&	410$\pm30$&	2\\	
&	1260.4221&	+24-64&	295$\pm10$&	2\\	
&	1304.3702&	-81.6&	63&	1\\
&	1304.3702&	-61&	68$\pm7$&	2\\	
&	1526.7066&	-50.5&	112$\pm12$&	2\\
Si\,III&	1206.5&	-74.3&	24&	1\\
&	1206.5&	-75&	33&	1\\
&	1206.5&	-103&	27&	1\\
&	1206.5&	-64-112&	398$\pm 12$&	2\\
Si\,IV&	1393.755&	-18-50&	186$\pm 13$&	2\\
&	1403&	+26-64&	167&	2\\	
S\,II&  1250.586&       -51 &   7$\pm 2$ & *\\
&       1253.805&             -49 &   17$\pm3$ & * \\
&	1259.518&	-77.0&	8.3&	1\\
&	1259.518&	-104.6&	$<$2.5&	1\\
&	1259.518&	-52&	7$\pm4$&	2\\	
S\,III&	1190&	-75.6	&$<$3.3&	1\\
Ar\,I&	1048&  ?&	$<$9.3&	1\\	
&	1067	&?&	$<$8.0&	1\\	
Fe\,II&	1055&	-81.6&	$<$3.2&	1\\
&	1097&	-78.6&	3.0&	1\\	
&	1145&	-78.6&	23.9&	1\\
&	1145&	-79&	21.1&	1\\	
&	1608&	20-50&	253$\pm21$&	2\\
\enddata

\tablenotetext{a}{An asterisk in the ``$\lambda$'' column
indicates absorption from an excited level.}
\tablenotetext{b}{References:
1 --- Shull (1977),
2 --- Howarth \&\ Phillips (1986),
3 --- Howarth \&\ Schmutz (1995),
4 --- This work}
\end{deluxetable}
\normalsize

\newpage


\scriptsize
\begin{deluxetable}{cllllll}
\tablecaption{Time-variability of P~Cygni lines}
\tablewidth{0pt}
\tablehead{
\colhead{Observation root name} & \colhead{Start time} &
\colhead{End time} & \colhead{EW$_{\rm edge}$\tablenotemark{a}} &
\colhead{EW$_{\rm abs}$} &
\colhead{EW$_{\rm em}$} & \colhead{$\lambda_{\rm c}$\tablenotemark{b}}}
\startdata
z19d0104t[1]& 00:28:38 & 00:31:00 & 2.61$\pm0.01$ & 9.41$\pm$0.01&
	-21.78$\pm$0.01&1239.02\\ 

z19d0104t[2] & 00:31:32 & 00:33:54 & 2.62 & 9.43 & -21.34 & 1238.65\\
z19d0104t[3] & 00:34:26 & 00:36:48 & 2.67 & 9.44 & -21.10 & 1238.29\\
z19d0104t[4] & 00:37:20 & 00:39:42 & 2.64 & 9.46 & -20.89 & 1237.92\\
z19d0105t[1] & 00:41:41 & 00:44:03 & 2.59 & 9.43 & -21.00 & 1240.49\\ 
z19d0105t[2] & 01:33:51 & 01:36:45 & 2.61 & 9.47 & -20.18 & 1240.12\\
z19d0105t[3] & 01:37:17 & 01:39:39 & 2.64 & 9.47 & -20.52 & 1239.76\\
z19d0105t[4] & 01:40:11 & 01:42:33 & 2.61 & 9.48 & -20.54 & 1239.39\\
z19d0106t[1] & 01:44:20 & 01:46:42 & 2.58 & 9.46 & -20.56 & 1240.50\\
z19d0106t[2] & 01:47:14 & 01:49:37 & 2.60 & 9.47 & -20.42 & 1240.14\\
z19d0106t[3] & 01:50:09 & 01:52:31 & 2.65 & 9.50 & -20.32 & 1239.77\\
z19d0106t[4] & 01:53:03 & 01:55:21 & 2.62 & 9.51 & -20.11 & 1239.40\\
z19d0107t[1] & 01:57:08 & 01:58:35 & 4.94 & 9.41   & -22.57  &  1548.37\\ 
z19d0107t[2] & 01:59:09 & 02:00:36 & 4.93 & 9.44   & -21.19  &  1548.01\\
z19d0107t[3] & 02:01:10 & 02:02:35 & 5.00 & 9.42   & -21.84  & 1547.65\\
z19d0107t[4] & 02:03:11 & 02:04:34 & 4.95 & 9.47   & -21.49  & 1547.3\\
z19d0108t[1] & 02:06:20 & 02:07:47 & 4.98 & 9.45   & -21.71  & 1549.79\\
z19d0108t[2] & 02:08:20 & 02:09:47 & 4.96 & 9.49   & -21.27 & 1549.43\\
z19d0108t[3] & 02:10:20 & 02:11:47 & 5.02 & 9.49   & -21.01 & 1549.08\\
z19d0108t[4] & 02:12:18 & 02:13:45 & 4.97 & 9.51   & -20.81 & 1548.72\\
z19d0109m[1] & 02:15:38 & 02:17:06 & 4.98& 9.47 & -21.34 & 1551.22\\
z19d0109m[2] & 02:17:38 & 02:19:06 & 4.95& 9.48 & -21.04 & 1550.86\\
z19d0109m[3] & 02:17:38 & 02:21:05 & 5.01& 9.48& -20.95  & 1550.50\\
z19d0109m[4] & 03:11:24 & 03:12:52 & 4.94& 9.46& -22.26 & 1550.15\\
\enddata
\tablenotetext{a}{EW$_{\rm edge}$ is the equivalent width of the
violet edge of the
absorption (all absorption shortward of saturation).  The equivalent
widths are relative to our choice of continuum and wavelength ranges
and show variability in emission with a weaker
in absorption.}
\tablenotetext{b}{$\lambda_{\rm c}$ is the central wavelength of the
exposure.}
\end{deluxetable}
\normalsize

 \newpage

\begin{deluxetable}{lcclcccc}
\tablecaption{Comparison Between Observed and Modeled Column Densities}
 \tablehead{
\colhead{Model\tablenotemark{a}} &
\colhead{$V_{\rm shell}$\tablenotemark{b}} & \colhead{$v_{\rm wind}$} &
\colhead{D$_{\rm kpc}$} & \colhead{N$_{\rm C\,IV}$\tablenotemark{c}} &
\colhead{N$_{\rm N\,V}$} & \colhead{N$_{\rm O\,VI}$}\\
\colhead{(Component)} & \colhead{($V_{\rm observed}$)}
& \colhead{} 
& \colhead{} 
& \colhead{} 
& \colhead{} 
& \colhead{} 
}
\startdata

(High velocity) & $\sim$75 &   & &           & 2.5(13)   &    \\
(Low velocity)  & $\lae$40  &   & & 1.5(14)   &  5(12) & $<$ 3.5(13)\\

A & 40  & 1700   &  2    & 5.0(12) & 4.0(12) & 2.5(13)  \\
B & 40  & 2800   &  2    & 5.8(12) & 4.8(12) & 2.9(13) \\
C & 80  & 1700   &  2    & 2.5(12) & 2.0(12) & 1.2(13)  \\
D & 80  & 2800   &  2    & 2.9(12) & 2.4(12) & 1.4(13)  \\
E & 80  & 1700   &  1    & 3.0(12) & 2.4(12) & 1.4(13) \\
F & 80  & 2800   &  1    & 3.5(12) & 2.9(12) & 1.7(13) \\ 

\enddata
\tablenotetext{a}{
Notes---For all models, the fractional abundances of carbon, 
nitrogen, and oxygen are given by $12+\log_{10} Y_{\rm C}=8.50$, 
$12+\log_{10} Y_{\rm N}=8.28$, and $12+\log_{10} Y_{\rm O}=8.54$.
The observationally determined mass-loss rate of HD~50896 has been
scaled with distance.}
\tablenotetext{b}{km s$^{-1}$}
\tablenotetext{c}{cm$^{-2}$}
\end{deluxetable}

\newpage
\section*{FIGURE CAPTIONS}

F{\sc ig}. 1: {\it HRS} spectra of HD~50896.  In (a), the region
around N\,V$\lambda$1240 is displayed.  (b) shows the spectrum near 
C\,IV$\lambda$1550.  Superimposed on the P~Cygni lines are the 
following narrow absorption lines in order of increasing wavelength: (a) 
N\,V$\lambda$1238.8 (slope discontinuity in P~Cygni profile), 
Mg\,II$\lambda\lambda$1238.9,1240.4,
N\,V$\lambda$1242.8, and S\,II$\lambda\lambda$1250.578,1253.805;
(b) C\,IV$\lambda\lambda$1548.2,1550.77,
C\,I,I$^{*}$,I$^{**} \lambda\lambda\lambda$1560,1560.3,1560.6.

F{\sc ig}. 2: The spectra deconvolved with the method of Jansson
(1984) and the known pre-COSTAR line spread function of the Large Science
Aperture.  The smooth curve shows our polynomial fit to the continuum, used
to calculate optical depths in the lines and as an upper bound for
deconvolution.  The raw (shallow absorption) and deconvolved lines
are superimposed.  In (a), a region around N\,V$\lambda$1240 is shown; 
(b) shows a region of C\,IV$\lambda$1550.

F{\sc ig}. 3: Optical depths $\tau (v)$ determined from doublets.
The square and triangle symbols represent, respectively,  
$\tau_{\rm blue}(v)$ 
and $2 \tau_{\rm red}(v)$. (a)---N\,V$\lambda\lambda 1238.8,1242.8$, 
(b)---Mg\,II$\lambda\lambda 1239.9,1240.4$, 
(c)--S\,II$\lambda\lambda 1250,1253$,
(d)---C\,IV$\lambda\lambda$1548.2,1550.8. 
For plots (a) and (d), we include theoretical profiles from our Model~E
based on the interstellar bubble theory.  We model the data in plots (b)
and (c) with a Gaussian with $b = 15$ km~s$^{-1}$.

F{\sc ig}. 4: The ratio $\tau_{\rm C\,IV}(v)/\tau_{\rm N\,V}(v)$.

F{\sc ig}. 5: The velocities of the absorption lines
observed with the {\it GHRS}.  Towards the top of the panel,
we show the velocities of N\,V, Mg\,II, S\,II, C\,IV, and C\,I.  The N\,V, 
Mg\,II, and S\,II  
lines were observed in the
same exposure, and have all been given a 15~km~s$^{-1}$ shift,
as discussed in the text.  The velocities of these lines are marked
with vertical lines which extend slightly further towards the top of the
panel than the C\,IV and C\,I lines, which extend slightly further
towards the bottom of the panel.  In the lower region of the panel,
we show the velocities expected for HD~50896, the expansion of S~308,
and a putative SNR along the line of sight.

F{\sc ig}. 6: The dotted curve shows the C\,I, C\,I$^{*}$, and
C\,I$^{**}$ lines observed by the HRS.  The jagged solid curve shows
our model based on the Na\,I absorption components previously
observed at higher resolution, and the shallow solid curve shows 
the model convolved with the HRS line spread function.


F{\sc ig}. 7: Electron density versus temperature, deduced from 
the C\,I, C\,I$^{*}$, C\,I$^{**}$ lines.  The solid lines show the 90\% 
confidence interval of electron density implied by the C\,I$^{*}$/C\,I 
ratio, the 
dashed line shows the result for the C\,I$^{**}$/C\,I ratio, and 
the dotted and dashed line shows the result for the C\,I$^{**}$/C\,I$^{*}$ 
ratio.


\newpage
\section{FIGURES}

\noindent {\bf Figure 1}

\vfill

\centerline{\psfig{figure=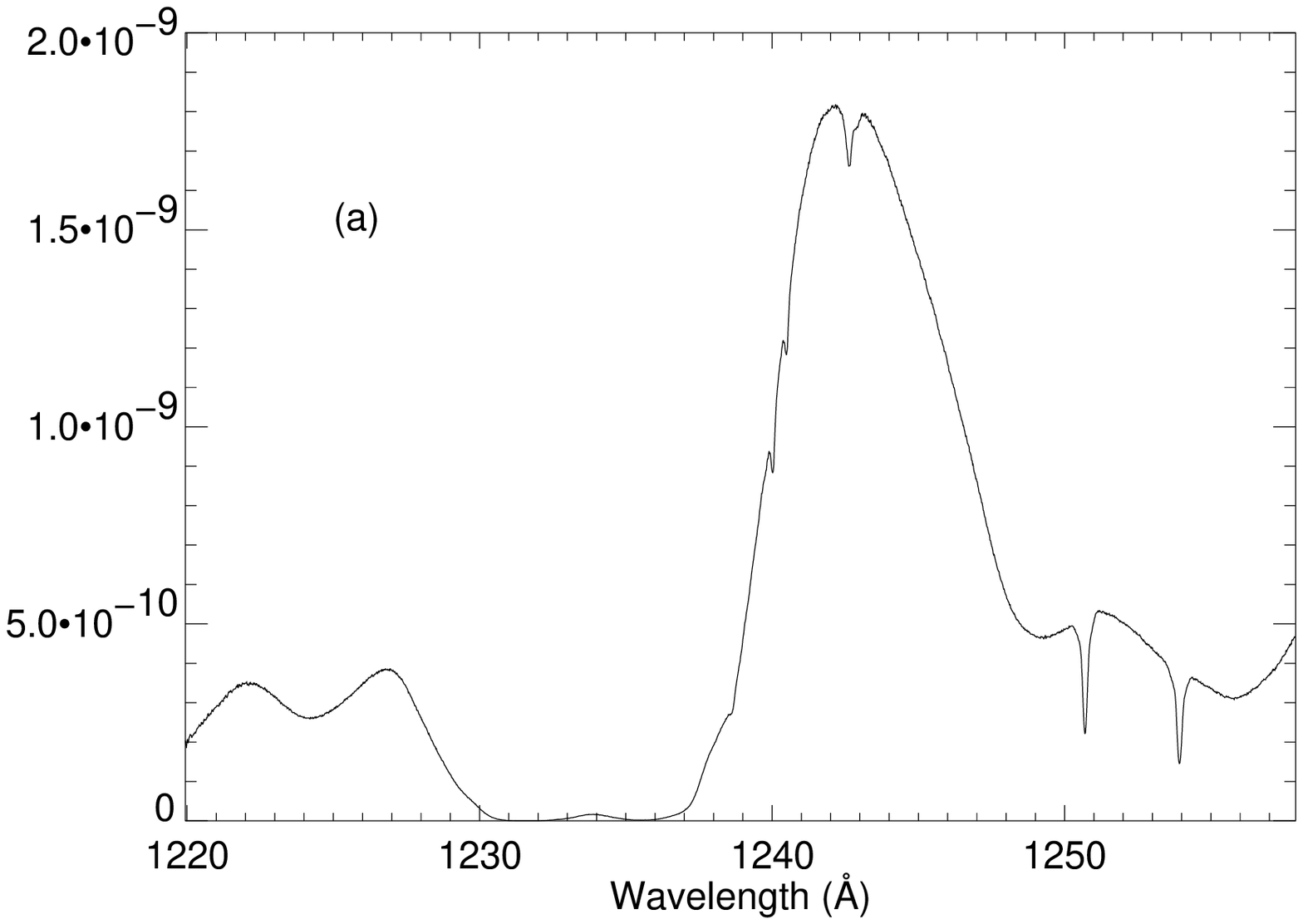,height=3.5in,width=5in}}

\vfill

\centerline{\psfig{figure=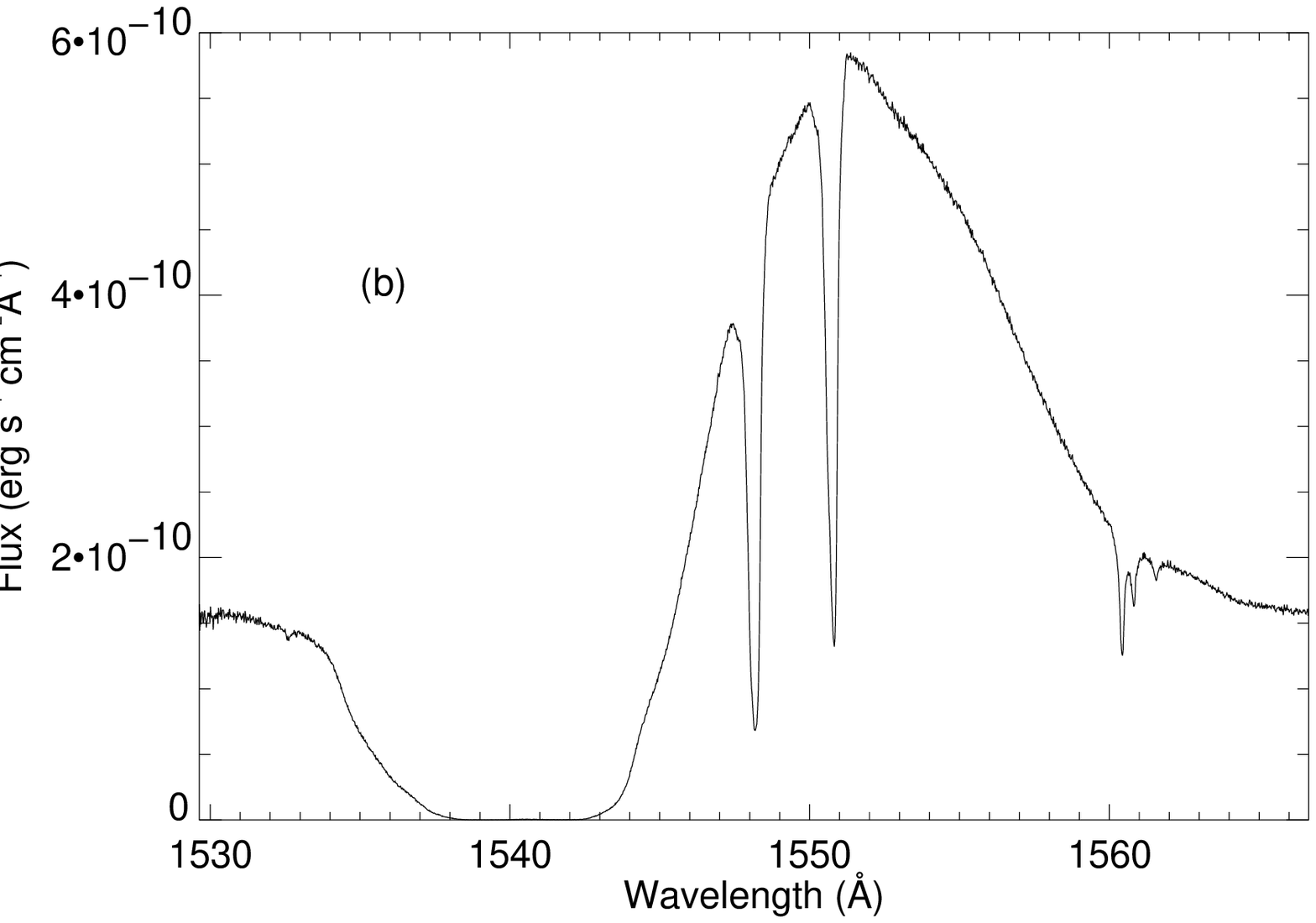,height=3.5in,width=5in}}

\vfill

\newpage
\noindent{\bf Figure 2}

\vfill

\centerline{\psfig{figure=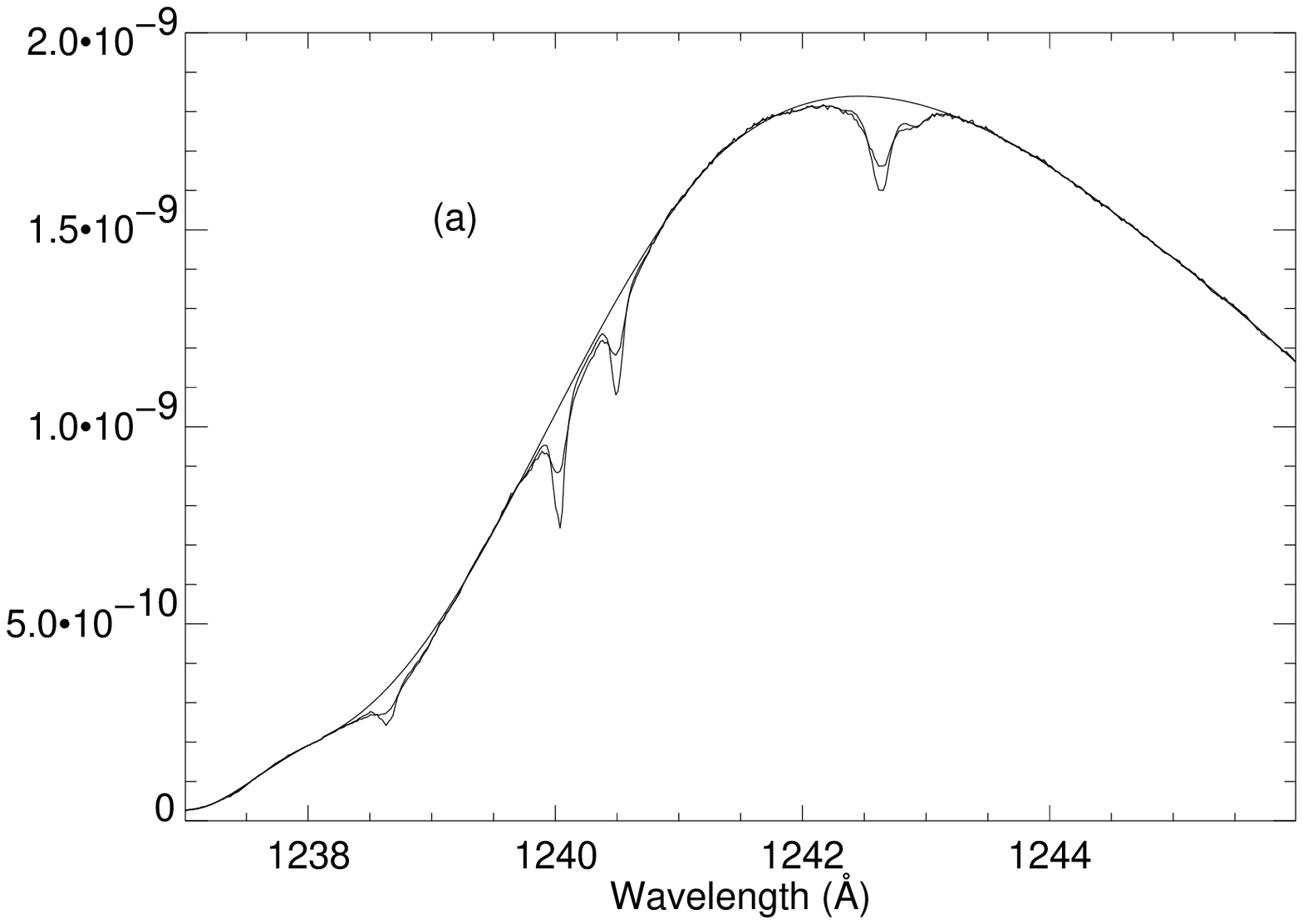,height=3.5in,width=5in}}
\vfill

\centerline{\psfig{figure=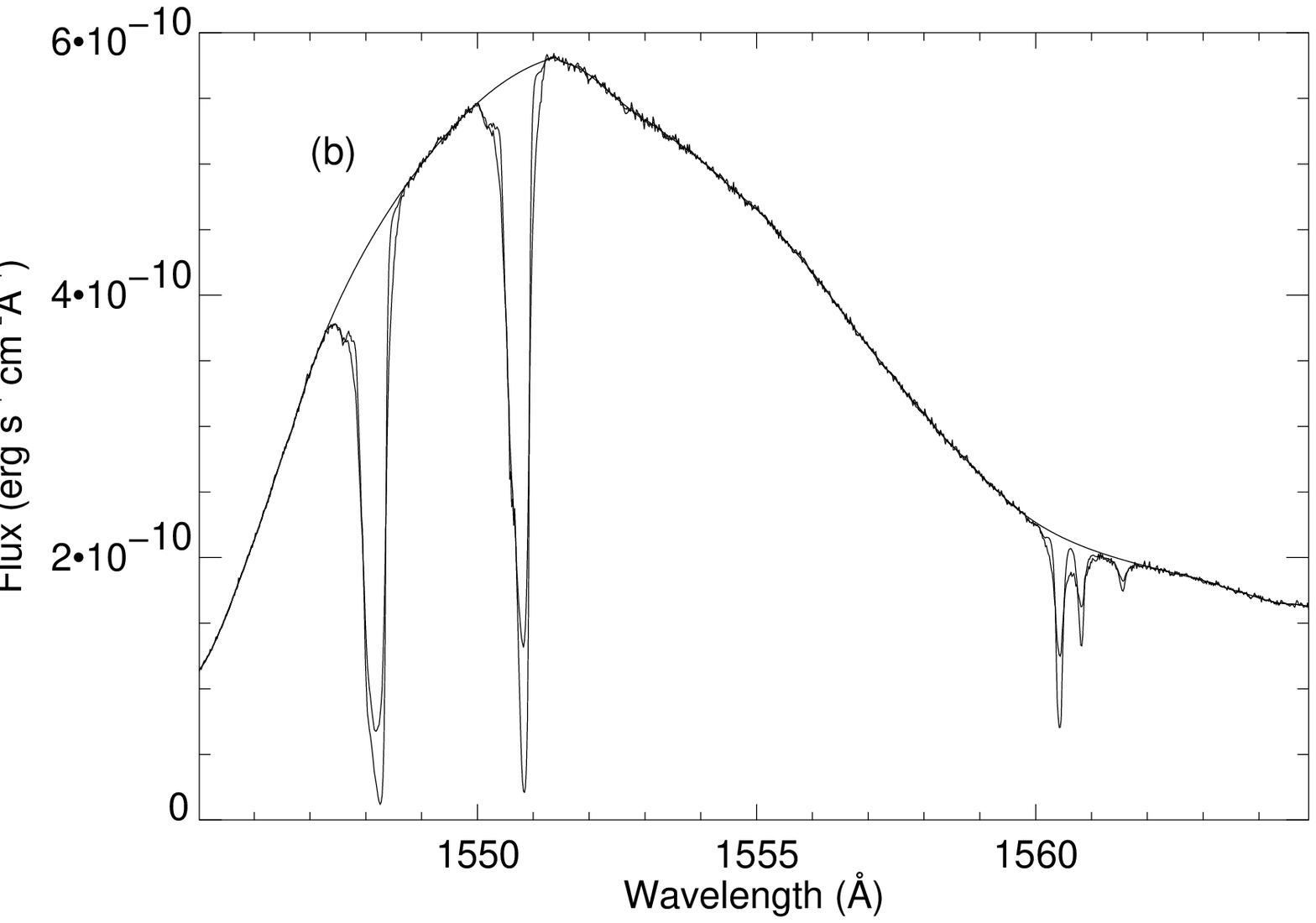,height=3.5in,width=5in}}

\vfill

\newpage

\noindent{\bf Figure 3}

\vfill

\centerline{\hbox{\psfig{figure=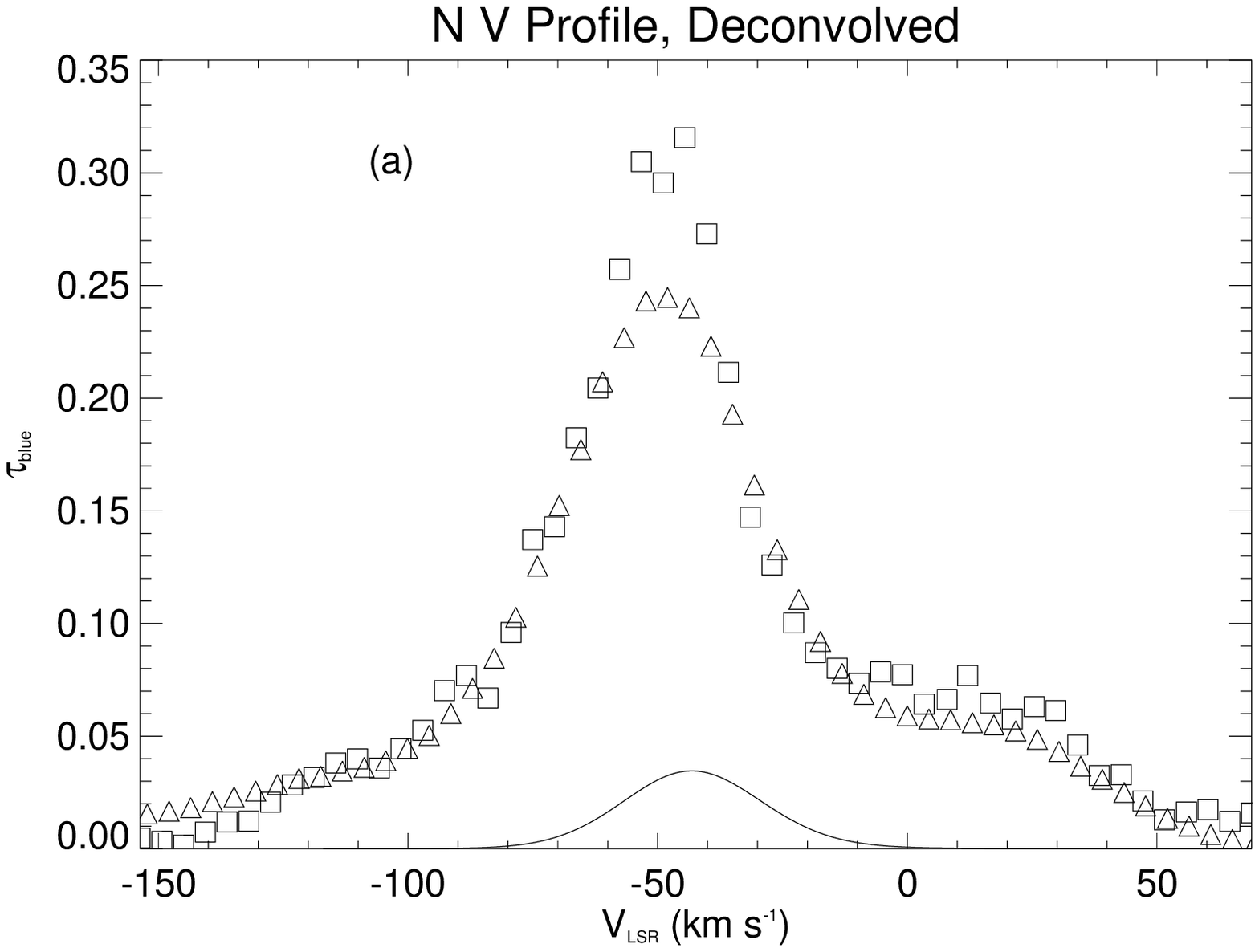,height=3.5in,width=3.6in}
\psfig{figure=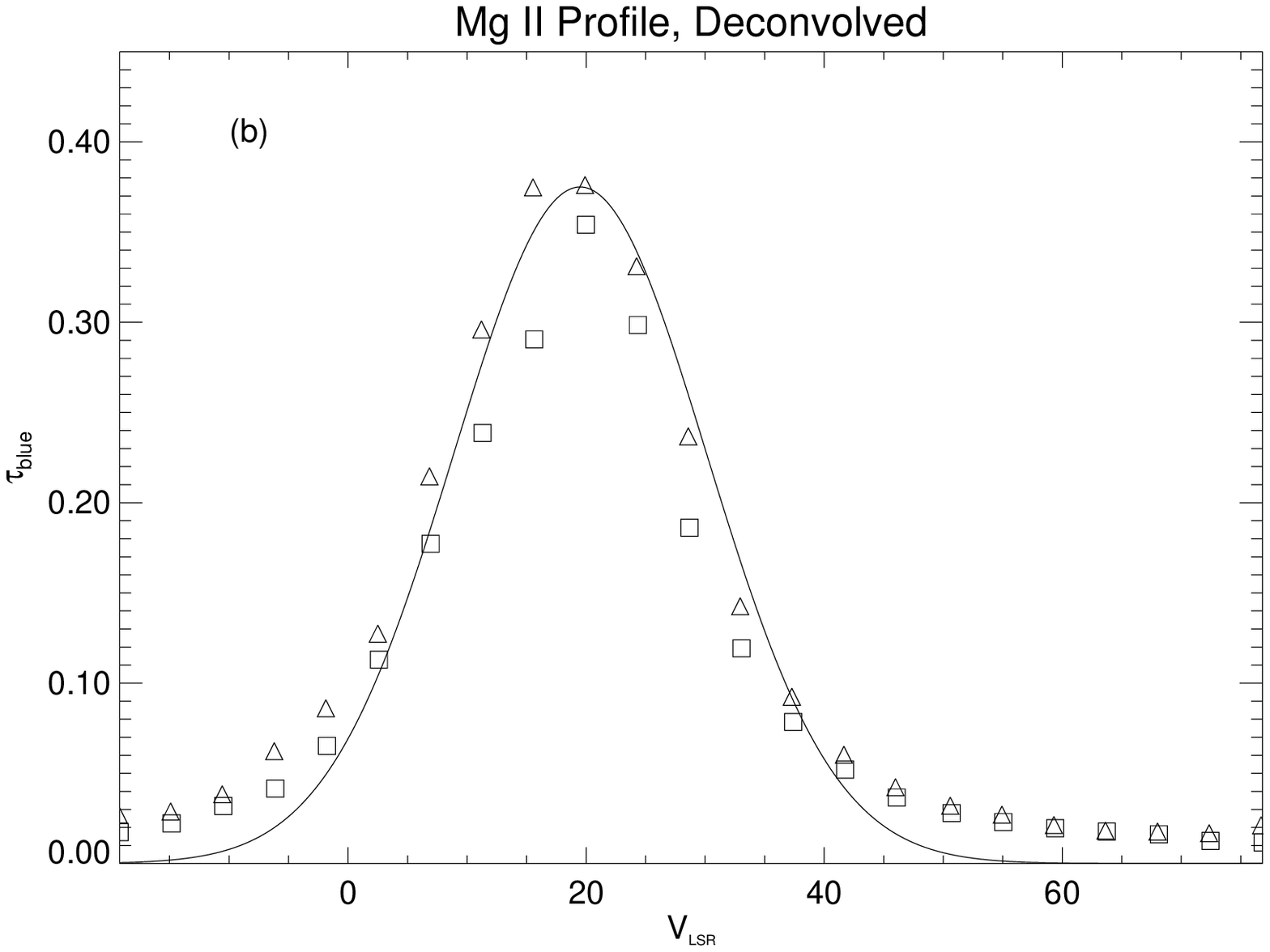,height=3.5in,width=3.6in}}}

\vfill

\centerline{\hbox{\psfig{figure=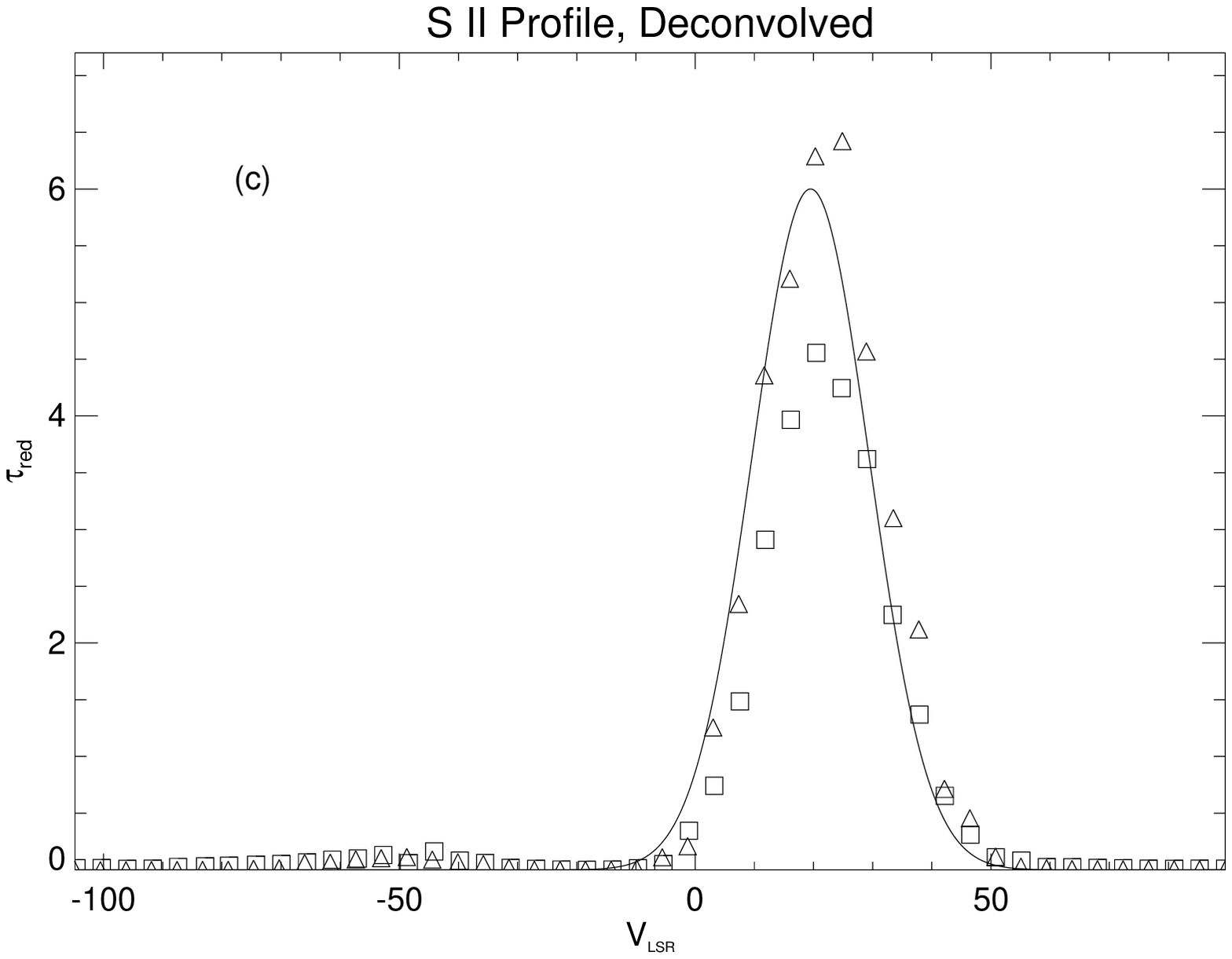,height=3.5in,width=3.6in}
\psfig{figure=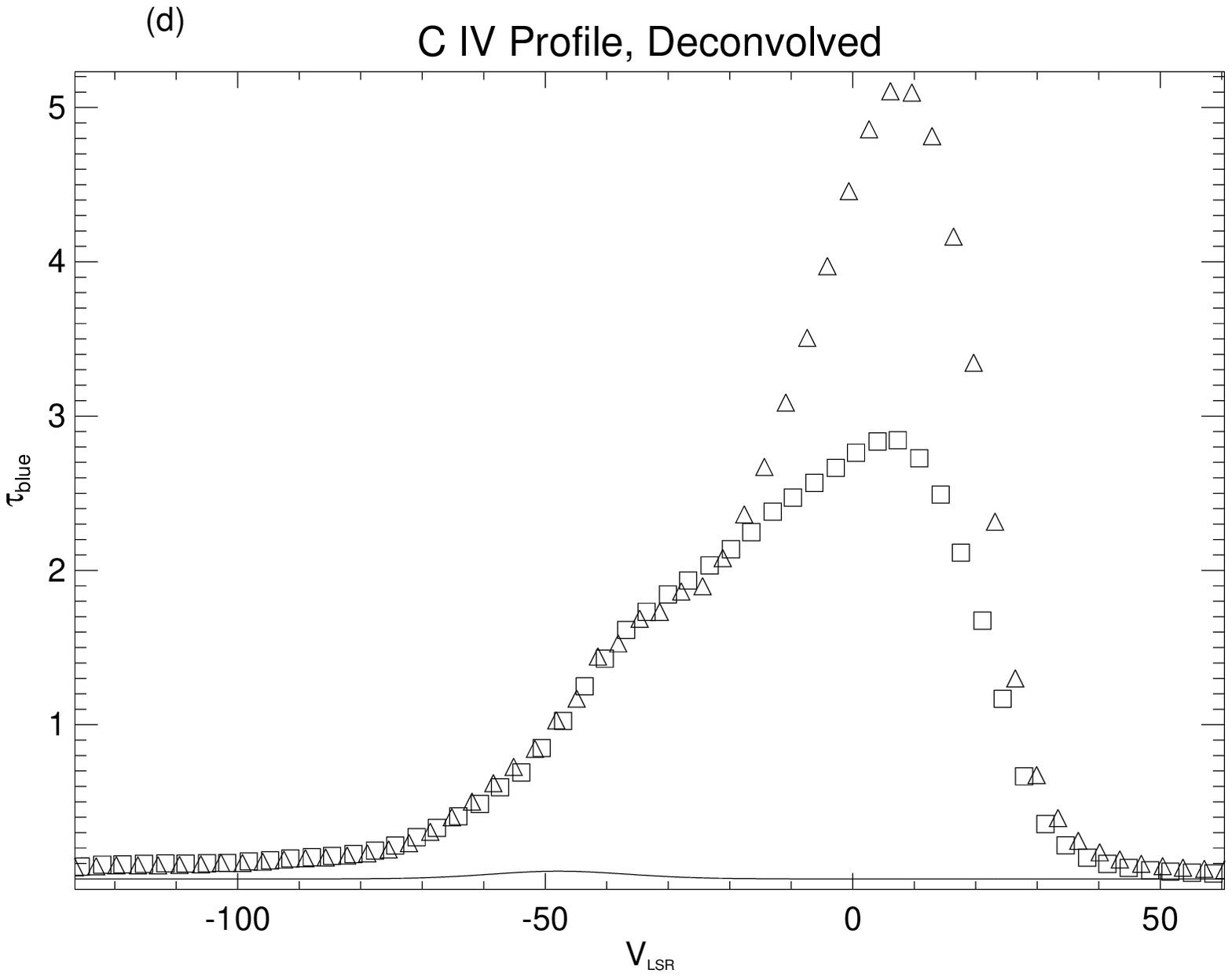,height=3.5in,width=3.6in}}}

\vfill

\newpage

\noindent{\bf Figure 4}

\vfill

\centerline{\hbox{\psfig{figure=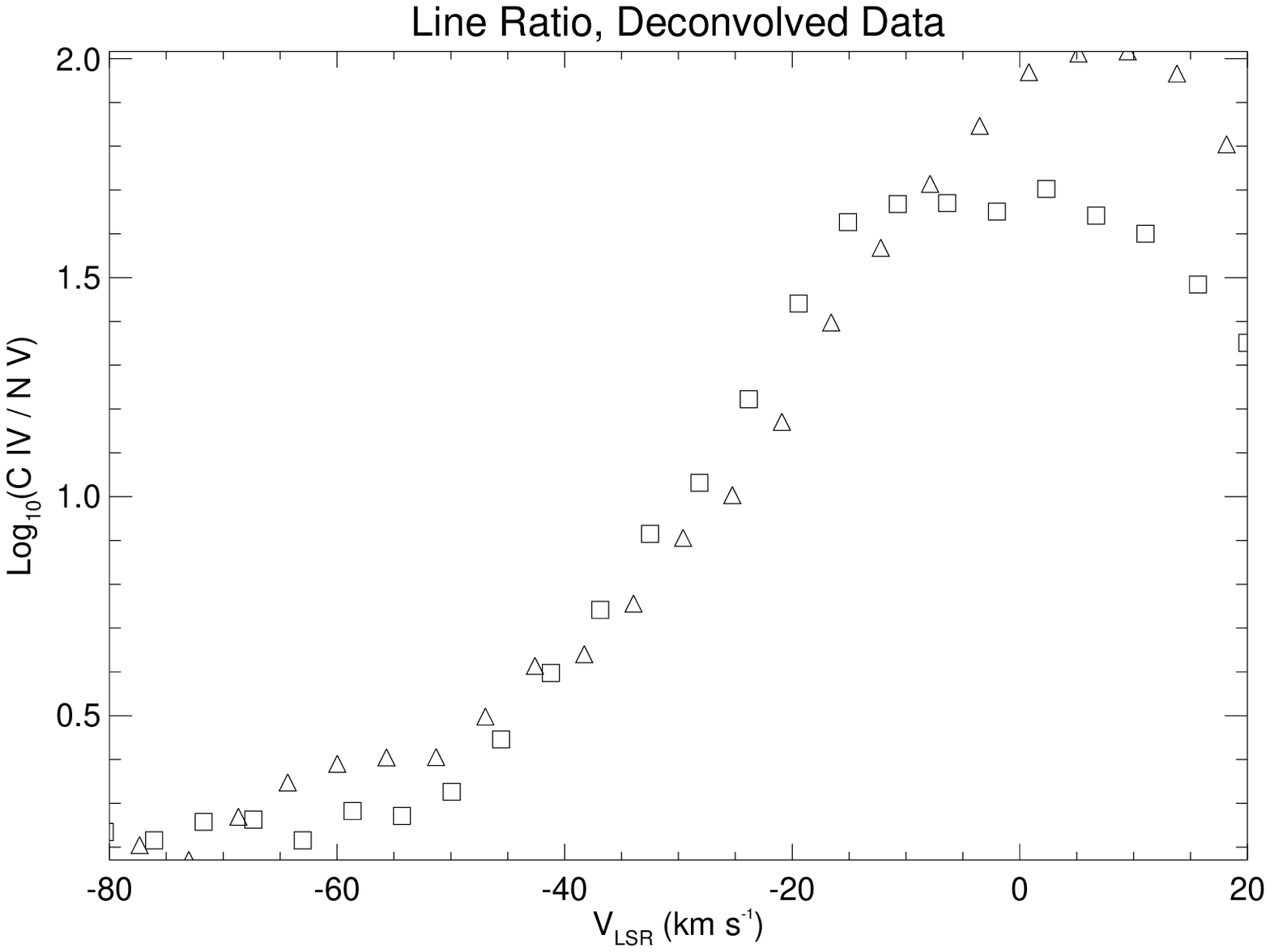,height=6in,width=7in}}}

\vfill

\newpage

\noindent{\bf Figure 5}

\centerline{\hbox{\psfig{figure=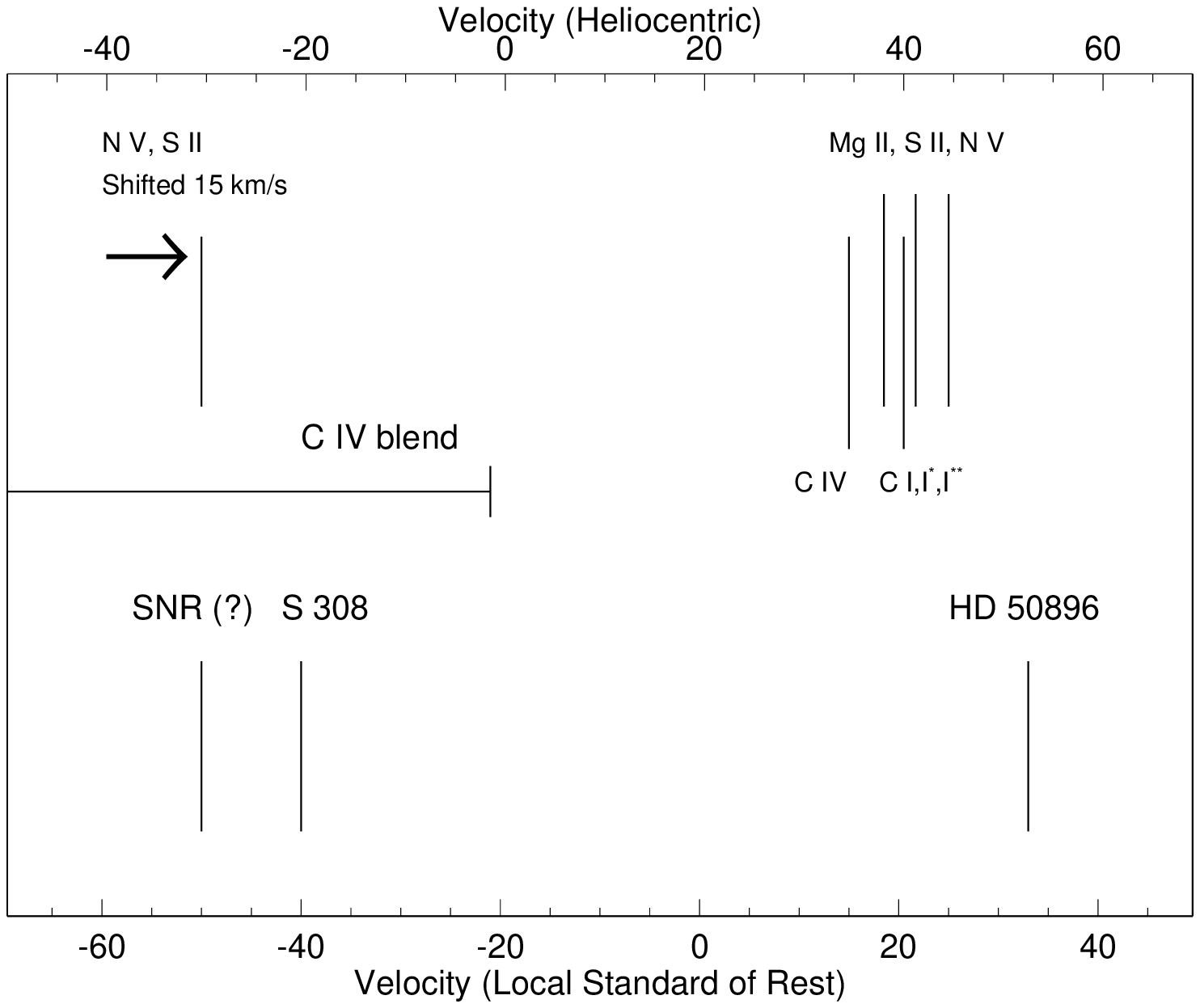,height=6in,width=7in}}}

\vfill

\newpage

\noindent{\bf Figure 6}

\vfill

\centerline{\hbox{\psfig{figure=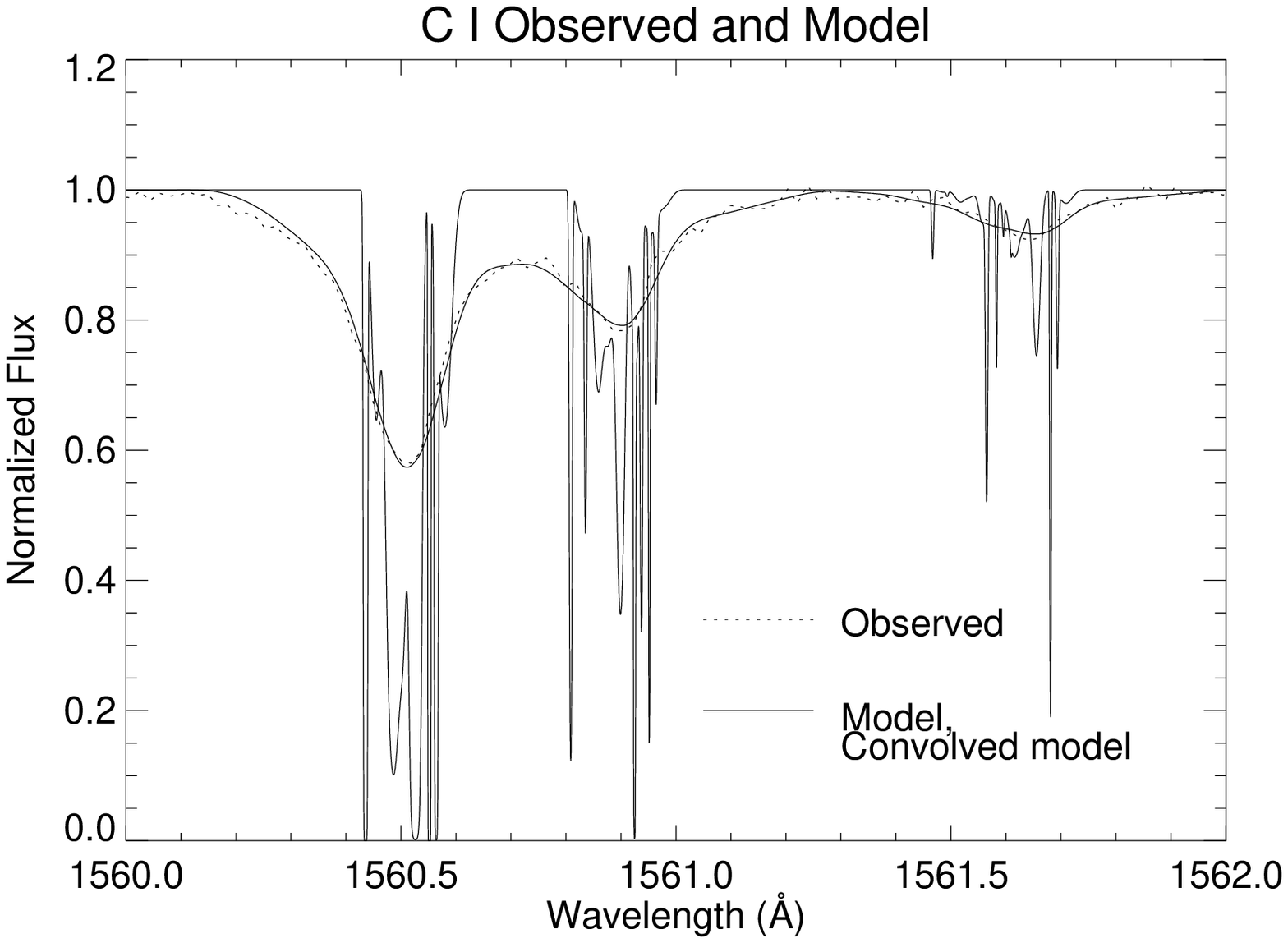,height=6in,width=7in}}}

\vfill

\newpage

\noindent{\bf Figure 7}

\centerline{\hbox{\psfig{figure=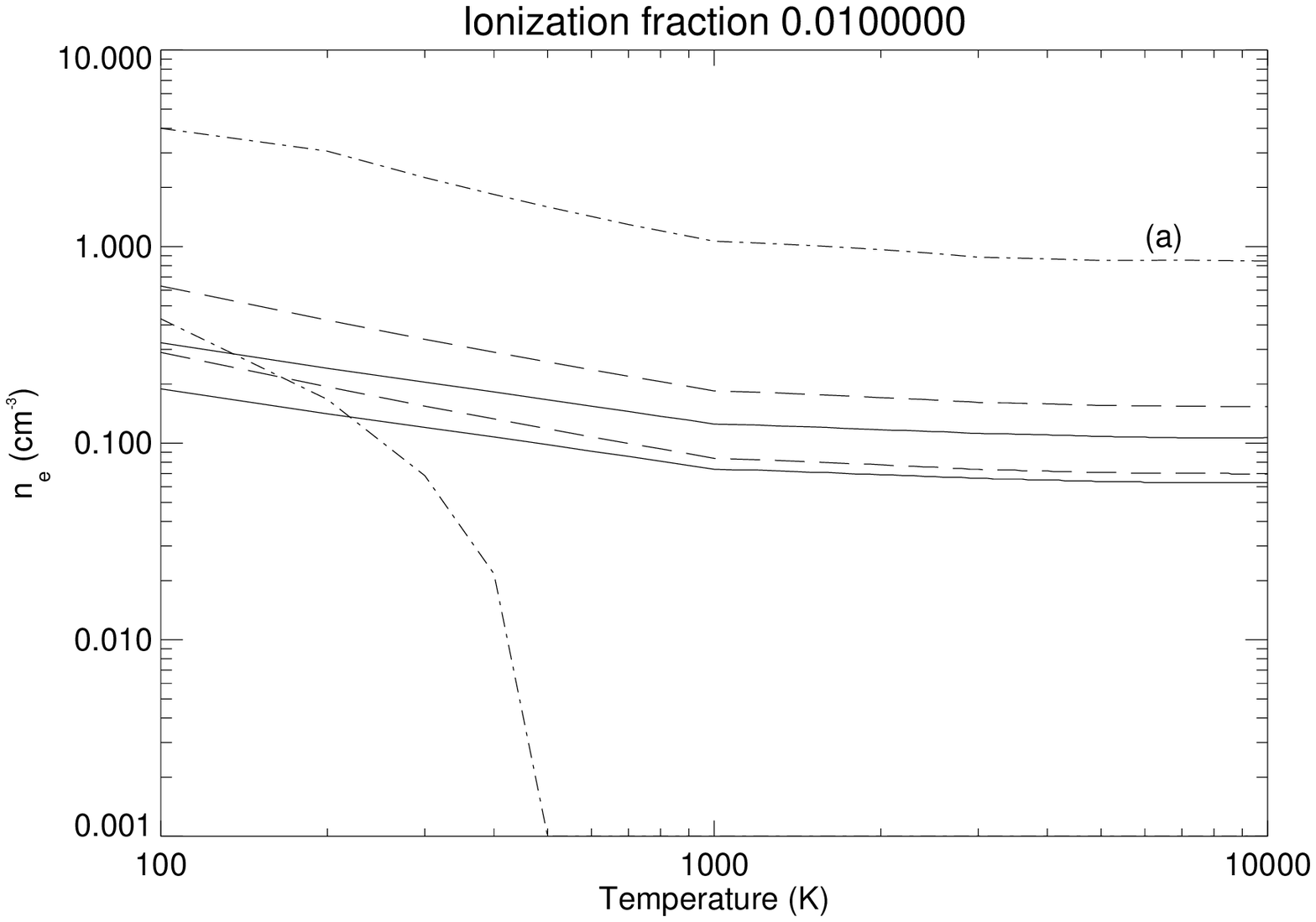,height=2.6in,width=5in}}}

\vfill

\centerline{\hbox{\psfig{figure=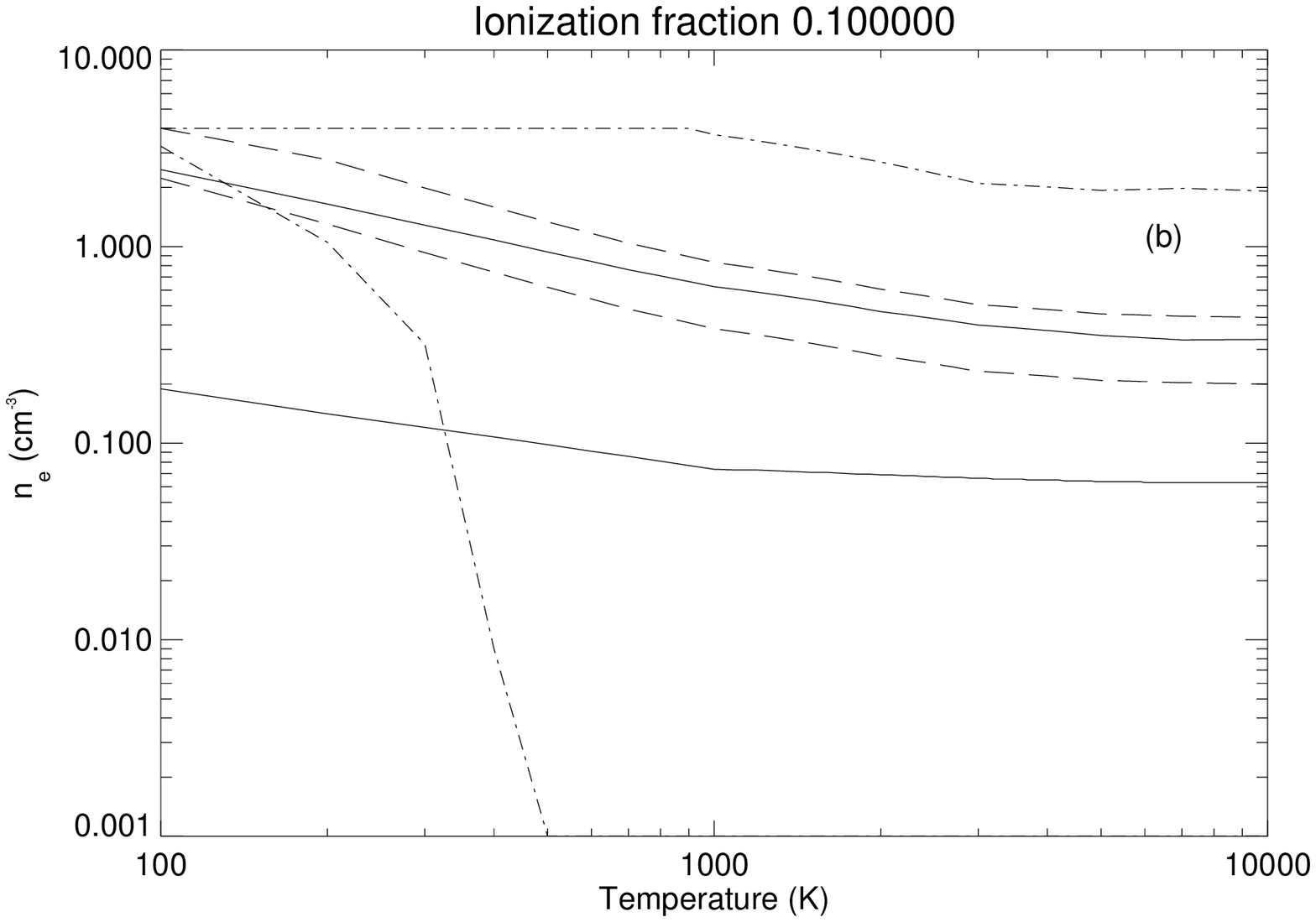,height=2.6in,width=5in}}}

\vfill

\centerline{\hbox{\psfig{figure=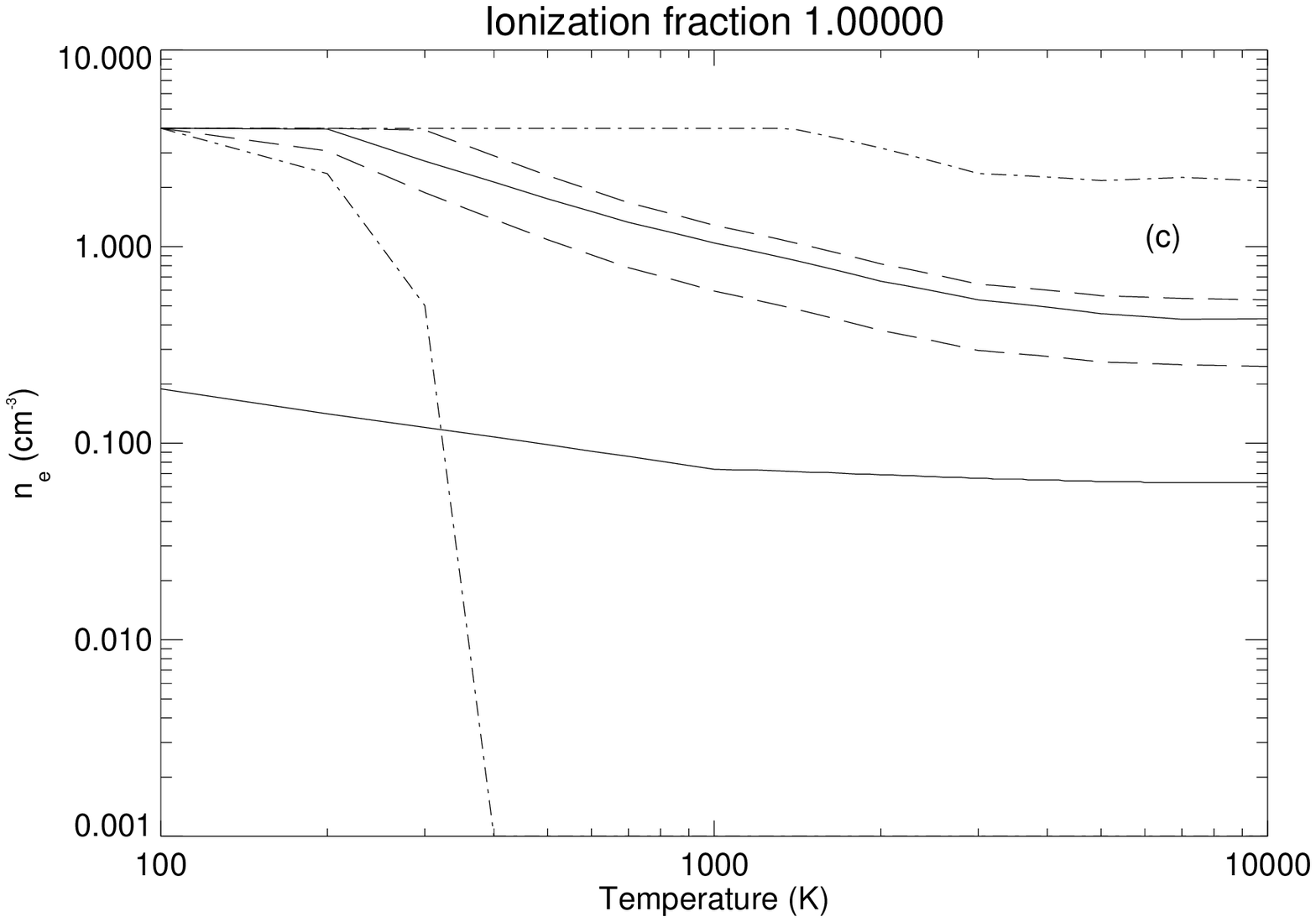,height=2.6in,width=5in}}}

\end{document}